\documentclass[twocolumn, trackchanges]{aastex61}

\newcommand{\project}[1]{\textsl{#1}}

\newcommand{\HST}{\project{HST}}
\newcommand{\Spitzer}{\project{Spitzer}}
\newcommand{\Kepler}{\project{Kepler}}

\usepackage{graphicx}

\newenvironment{my_itemize}{
    \begin{itemize}
              \setlength{\itemsep}{1pt}
                    \setlength{\parskip}{0pt}
                      \setlength{\parsep}{0pt}}{\end{itemize}
                      }

\shorttitle{Phase Curves of WASP-103\lowercase{b}}
\shortauthors{Kreidberg et al.}

\begin{document}

\title{Global Climate and Atmospheric Composition of the Ultra-Hot Jupiter WASP-103\lowercase{b} from \HST\ and \Spitzer\ Phase Curve Observations}


\author{Laura Kreidberg}
\affiliation{Harvard-Smithsonian Center for Astrophysics, 60 Garden Street, Cambridge, MA 02138}
\affiliation{Harvard Society of Fellows, 78 Mount Auburn Street, Cambridge, MA 02138}

\author{Michael R. Line}
\affiliation{School of Earth \& Space Exploration, Arizona State University, Tempe AZ 85287, USA}

\author{Vivien Parmentier}
\affiliation{Aix Marseille Univ, CNRS, LAM, Laboratoire d'Astrophysique de Marseille, Marseille, France}

\author{Kevin B. Stevenson}
\affiliation{Space Telescope Science Institute, 3700 San Martin Drive, Baltimore, MD 21218, USA}

\author{Tom Louden}
\affiliation{Department of Physics, University of Warwick, Coventry, CV4 7AL, UK}

\author{Mick\"{a}el Bonnefoy}
\affiliation{Univ. Grenoble Alpes, IPAG, F-38000 Grenoble, France. CNRS, IPAG, F-38000 Grenoble, France}

\author{Jacqueline K. Faherty}
\affiliation{American Museum of Natural History, Department of Astrophysics, Central Park West at 79th Street, New York, NY 10034, USA}

\author{Gregory W. Henry}
\affiliation{Center of Excellence in Information Systems, Tennessee State University, Nashville, TN 37209, USA}

\author{Michael H. Williamson}
\affiliation{Center of Excellence in Information Systems, Tennessee State University, Nashville, TN 37209, USA}

\author{Keivan Stassun}
\affiliation{Vanderbilt University, Dept. of Physics and Astronomy, 6301 Stevenson Center Ln, Nashville TN, 37235, USA}

\author{Thomas G.\ Beatty}
\affiliation{Department of Astronomy \& Astrophysics, The Pennsylvania State University, 525 Davey Lab, University Park, PA 16802}
\affiliation{Center for Exoplanets and Habitable Worlds, The Pennsylvania State University, 525 Davey Lab, University Park, PA 16802}

\author{Jacob L. Bean}
\affiliation{Department of Astronomy \& Astrophysics, University of Chicago, 5640 S. Ellis Avenue, Chicago, IL 60637, USA}

\author{Jonathan J. Fortney}
\affiliation{Department of Astronomy and Astrophysics, University of California, Santa Cruz, CA 95064}

\author{Adam P. Showman}
\affiliation{Department of Planetary Sciences and Lunar and Planetary Laboratory, University of Arizona, Tucson, Arizona 85721, USA
}

\author{Jean-Michel D\'{e}sert}
\affiliation{Anton Pannekoek Institute for Astronomy, University of Amsterdam, Science Park 904, 1098 XH Amsterdam, The Netherlands}

\author{Jacob Arcangeli}
\affiliation{Anton Pannekoek Institute for Astronomy, University of Amsterdam, Science Park 904, 1098 XH Amsterdam, The Netherlands}

\begin{abstract}
We present thermal phase curve measurements for the hot Jupiter WASP-103b observed with \emph{Hubble}/WFC3 and \emph{Spitzer}/IRAC.  The phase curves have large amplitudes and negligible hotspot offsets, indicative of poor heat redistribution to the nightside.  We fit the phase variation with a range of climate maps and find that a spherical harmonics model generally provides the best fit. The phase-resolved spectra are consistent with blackbodies in the WFC3 bandpass, with brightness temperatures ranging from $1880\pm40$ K on the nightside to $2930 \pm 40$ K on the dayside. The dayside spectrum has a significantly higher brightness temperature in the \Spitzer\ bands, likely due to CO emission and a thermal inversion.  The inversion is not present on the nightside. We retrieved the atmospheric composition and found the composition is moderately metal-enriched ($\mathrm{[M/H]} = 23^{+29}_{-13}\times$ solar) and the carbon-to-oxygen ratio is below 0.9 at $3\,\sigma$ confidence. In contrast to cooler hot Jupiters, we do not detect spectral features from water, which we attribute to partial H$_2$O dissociation.  We compare the phase curves to 3D general circulation models and find magnetic drag effects are needed to match the data.  We also compare the WASP-103b spectra to brown dwarfs and young directly imaged companions and find these objects have significantly larger water features, indicating that surface gravity and irradiation environment play an important role in shaping the spectra of hot Jupiters. These results highlight the 3D structure of exoplanet atmospheres and illustrate the importance of phase curve observations for understanding their complex chemistry and physics.
\end{abstract}

\keywords{planets and satellites: individual (WASP-103b), planets and satellites: atmospheres}

\section{Introduction} \label{sec:intro}
Planets are round, rotating, and irradiated on one hemisphere at a time -- all of which contribute to rich spatial structure in their climate and atmospheric composition.  Short period, tidally locked planets are an extreme example, with one hot, continuously illuminated side. This asymmetry is expected to produce large gradients in temperature, chemistry, and cloud coverage with longitude \citep{showman09, kataria16, parmentier16}, and provides an opportunity to learn about atmospheric dynamics in a very different regime from the Solar System planets.

Exoplanets are so distant that they are generally not spatially resolved from their host stars, but it is still possible to reveal inhomogeneities in their atmospheres by observing the total system flux. One approach is to measure a phase curve, which consists of continuous monitoring of the planet-to-star flux ratio over a complete orbital revolution of the planet. This observation is sensitive to different longitudes at each orbital phase of the planet.  The first phase curve of an exoplanet was observed with \Spitzer\ for the hot Jupiter $\nu$ Andromedae b by \cite{harrington06}, followed by additional \Spitzer\ observations for about a dozen more systems \citep[cataloged in][]{parmentier17}.  These observations revealed large day-night temperature contrasts (in excess of 300 K), and eastward shifted peak brightness due to heat circulation, as predicted by 3D models \citep{showman02}.  These infrared measurements were complemented by optical phase curves from \Kepler\ that show evidence for reflected light from patchy and possibly variable dayside clouds with a range of compositions \citep{borucki09, demory13, hu15, armstrong16, parmentier16}. A \emph{spectroscopic} phase curve was observed for WASP-43b with \emph{Hubble}/Wide Field Camera 3 (\HST/WFC3) in the near-infrared, which provided the first phase-resolved measurements of an exoplanet's water abundance and thermal structure \citep{stevenson14, stevenson17}.

In this paper we present spectroscopic phase curve observations of the hot
Jupiter WASP-103b, measured with \HST/WFC3 and Spitzer/IRAC. This planet is an
ideal target for phase curve observations, with an orbital period of just 22
hours and an equilbrium temperature of 2500 K.  WASP-103b is slightly
larger than Jupiter, with a mass and radius of $1.49\pm0.09\,M_\mathrm{Jup}$ and
$1.53^{+0.05}_{-0.07}\,R_\mathrm{Jup}$, respectively. The host star is a main-sequence F8 dwarf with an effective temperature of $6110 \pm 160\,$ K \citep{gillon14}. Previous observations of WASP-103b's atmosphere revealed a blackbody-like dayside emission spectrum, with possible evidence for a $K_\mathrm{S}$-band emission feature \citep{cartier17, delrez18}. The optical transmission spectrum shows evidence for sodium and potassium absorption features that are consistent with expectations for a cloud-free atmosphere \citep{lendl17}.

WASP-103b is an archetype of the class of ultra-hot Jupiters, with orbital
periods of about one day and dayside temperatures typically $>2000\,\mathrm{K}$.
These very hot planets were initially predicted to have inverted temperature
pressure profiles due to strong optical absorption by TiO/VO in the upper
atmospheres \citep{hubeny03, fortney08}; however, observations of their emission
spectra have been inconclusive on their thermal structure and composition. In
the near-infrared, where water is the dominant absorber, some spectra show water
absorption features, some show emission features, and some are consistent with
blackbody models \citep{madhusudhan11, crossfield12, stevenson14b, haynes15,
evans16, beatty17a, beatty17b, sheppard17, arcangeli18, mansfield18}.  A variety
of explanations have been proposed for these results, including low metallicity
or high carbon-to-oxygen compositions, dayside clouds, and finely tuned
isothermal temperature pressure profiles. Recently, \cite{arcangeli18,
lothringer18} showed that water dissociation and H- opacity on the hot dayside play an important role in the atmospheres of these ultra-hot planets and may be responsible for some of the blackbody-like near-IR spectra. In this work, we put these results in context by investigating the global thermal structure and composition of the ultra-hot Jupiter WASP-103b.

The structure of the paper is the following: in \S\,\ref{sec:observations} we describe the observations and data reduction. \S\,\ref{sec:fits} details the models fit to the phase curves. In \S\,\ref{sec:results} we discuss results, including the phase curve amplitudes and hotspot offsets, the phase-resolved spectra, estimates of the planet's climate, and the transmission spectrum. In \S\,\ref{sec:gcm} and \ref{sec:comparison}, we compare the observations to general circulation model (GCM) predictions and spectra from similar temperature stars and directly imaged companions. \S\,\ref{sec:summary} concludes.

\section{Observations and Data Reduction}
\label{sec:observations}
We observed two full-orbit phase curves of WASP-103b with \HST/WFC3 and one each with \Spitzer/IRAC at 3.6 and 4.5 $\mu$m (from HST Program 14050 and Spitzer Program 11099, PI: L. Kreidberg). We also reduced two \HST/WFC3 secondary eclipse observations of WASP-103b from \HST\ Program 13660 (PI: M. Zhao).

\subsection{\HST/WFC3}
The \HST\ phase curve observations consisted of two visits on 26-27 February and 2-3 August 2015. Each visit was 15 orbits in duration and spanned 23 hours. The last half of orbit 15 in each visit was used for a gyro bias update and does not produce useable science data.  We took a direct image of the star with the F126N filter at the beginning of each orbit to determine the wavelength solution zero-point. The remainder of the orbit consisted of time-series spectroscopy with the G141 grism ($1.1 - 1.7$ $\mu$m) and the 256 x 256 pixel subarray. We used the SPARS10/NSAMP = 15 read-out mode, which has an exposure time of 103 seconds. To optimize the duty cycle of the observations, we used the spatial scan observing mode with a scan rate of 0.03 arcsec/s, alternating between forward and backward scanning on the detector. The scan height was 25 pixels and the peak counts were $3.5\times10^4$ photoelectrons per pixel. We collected a total of 18 spatial scan exposures per orbit.  The two eclipse observations from Program 13660 had a similar observing setup \citep[described in detail in][]{cartier17}.  

We reduced the data from both programs using a custom pipeline developed for past analyses of WFC3 data \citep[for details see][]{kreidberg14a, kreidberg14b, kreidberg15b}. Briefly, we use the optimal extraction algorithm of \cite{horne86} to extract each up-the-ramp sample (or ``stripe") separately. The stripes are then summed to create the final spectrum. For each stripe, the extraction window is 24 pixels high and centered on the stripe midpoint. We estimate the background from the median of a region of the detector that is uncontaminated by the target spectrum (rows 5-50). The typical background counts are low (10-15 photoelectrons per pixel, roughly 0.03\% of the peak counts from the target star). We note that the extracted spectrum includes flux from a nearby companion star, which is separated from WASP-103 by less than two pixels \citep[0.2";][]{wollert15}. We account for this contamination later in the analysis. 

\subsection{\Spitzer}
We also obtained \Spitzer/IRAC observations with 3.6 and 4.5 $\mu$m photometric filters (referred to as Channel 1 and Channel 2, respectively). The observations had the following setup. Each phase curve observation consisted of 30 hours of time series photometry, beginning three hours prior to one secondary eclipse and ending three hours after a second eclipse.  We read out the full array and used 12 s exposures to maximize the duty cycle without saturating the detector. To minimize the intrapixel effect (variations in flux caused by imprecise pointing), we did not dither and also used PCRS peak-up\footnote{\url{http://irsa.ipac.caltech.edu/data/SPITZER/docs/irac/pcrs\_obs.shtml}} to improve the pointing accuracy. We began each observation with a 30-minute position settling period, followed by three Astronomical Observation Requests (AORs) of equal duration. At the beginning of each AOR, the telescope was repointed to position the target in the ``sweet spot" of the detector, where the response is fairly uniform over the pixel.

The data were reduced with the POET pipeline \citep{stevenson12, cubillos13}.
The pipeline starts by identifying and flagging bad pixels using a
double-iteration 4-sigma outlier rejection routine along the time axis.  This is
followed by performing 2D Gaussian centroiding on each frame, which is shown to
provide the most precise centers for Spitzer data \citep{Lust2014}.  The target
remains centered near the sweet spot for the entire AOR in each observation,
with a maximum drift of 0.1 pixels.  Next, POET uses sub-pixel
(5$\times$-interpolated) aperture photometry \citep{Harrington2007} to subtract
the background and sum the flux within a specified radius.  Chosen from a grid
of apertures between 2 and 4 pixels, we find that an aperture size of 2.75
pixels minimizes the residual noise in the light curve fits.  For the
background, we use an annulus with inner and outer radii of 7 and 15 pixels,
respectively.  The contaminating flux from the nearby star is within the same
pixel as the target, so we included it in the photometry and corrected it in the
light curve fits.  A similarly strategy has been applied to successfully analyze
dozens of Spitzer data sets \citep[e.g. ][]{Stevenson2010, Stevenson2012a,
Stevenson2012b, stevenson14, stevenson14b, Stevenson2016a, stevenson17,
Campo2011, Nymeyer2011, cubillos13, Blecic2013, Blecic2014, diamond-lowe14}.


\subsection{Photometric Monitoring}
To assess how stellar activity might impact the phase curve observations, we monitored WASP-103's photometric variability over 158 nights during 2014 - 2016 with the Tennessee State University Celestron 14-inch (C14) automated imaging telescope (AIT), located at Fairborn Observatory in southern Arizona \citep[][]{henry99}.  The observations of WASP-103 were made in the Cousins R passband with an SBIG STL-1001E CCD camera.  Each observation consisted of 4--10 consecutive exposures on WASP-103 along with several dozen comparison stars in the same field. The individual consecutive frames were co-added and reduced to differential magnitudes (i.e., WASP-103 minus the mean brightness of the six best comparison stars). The nightly observations were corrected for bias, flat-fielding, and differential atmospheric extinction.  For each season, we determined extinction corrections with a linear least-squares fit to nightly differential magnitude as a function of airmass.

The photometric analyses are summarized for each observing season in Table\,\ref{tab:photometry}.  The standard deviations of a single observation with respect to the corresponding seasonal means are given in column~4; the mean of the three standard deviations is 0.0058~mag, suggesting there is little night-to-night variation in WASP-103.  The three seasonal mean brightness values given in column 5 scatter about their grand mean with a standard deviation of 0.0036 mag, but we note that the most discrepant mean is from the third season, for which we have only partial coverage. Therefore, our results do not completely rule out low-level, year-to-year variability of $<0.001$~mag.

\begin{figure}
\includegraphics[width = 0.5\textwidth]{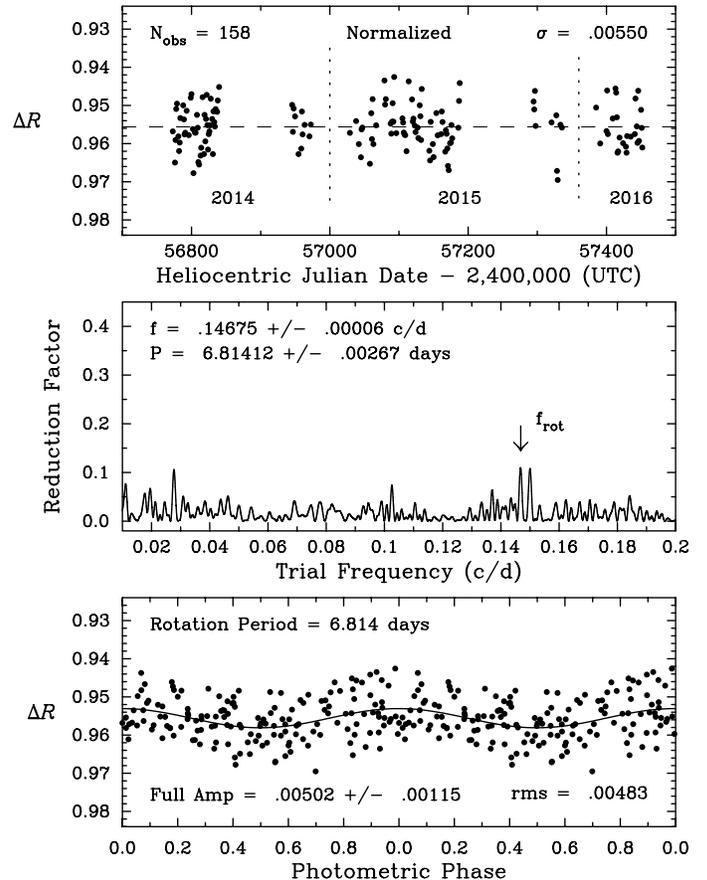}
\caption{$Top$: The normalized nightly Cousins $R$ band photometric dataset for WASP-103, acquired with the C14 automated imaging telescope at Fairborn Observatory. Vertical dashed lines denote separate observing seasons. Gaps are due to target visibility and the Arizona monsoon season (July - September). $Middle$: The frequency spectrum of the normalized dataset suggests low-amplitude variability with a period of 6.814~days. $Bottom$: The normalized dataset phased to the 6.814-day period, which we interpret as rotational modulation of a star spot or spots. A least-squares sine fit to the 6.814-day rotation period gives a peak-to-peak amplitude of just 0.005~mag.}
\label{fig:photometry}
\end{figure}

To maximize the possibility of detecting WASP-103's rotation, we normalized the
photometry such that each observing season has the same mean, thereby removing
any long-term variability in WASP-103 and/or the comparison stars
(Figure\,\ref{fig:photometry}, top panel).  To estimate the stellar rotation
period, we performed a periodogram analysis of the normalized data set based on
least-squares fitting of sine curves.  The resulting frequency spectrum and the
phase curve computed with the best period are shown in the middle and lower
panels of Figure\,\ref{fig:photometry} respectively. The best-fit period is
6.814 days, which agrees closely with the estimated stellar rotation period of
6.855 days \citep[based on the projected stellar rotation velocity and stellar
radius reported in][]{gillon14}. There are two nearby peaks in the
periodogram (panel b of Figure\,\ref{fig:photometry}) that are one-year aliases
of each other, and we chose the peak that better matches the stellar rotation velocity. The peak-to-peak variability amplitude is 0.005 mag.  Based on the formalism in \cite{zellem17}, we calculate that this variability will bias the measured eclipse depth by $\lesssim10$\,parts per million (ppm) from epoch to epoch, which is well below the photon-limited precision of our measurements.

\begin{deluxetable}{ccccc}
	\tablewidth{0pt}
	\tablecaption{Photometric Observations of WASP-103 \label{tab:photometry}}
	\tablehead{
		\colhead{Observing} & \colhead{$N_{obs}$} & \colhead{Date Range} &
	\colhead{Sigma} & \colhead{Seasonal Mean} \\
	\colhead{Season} &  \colhead{\,} & \colhead{(HJD-2,400,000)} & \colhead{(mag)} & \colhead{(mag)}}
	\startdata
	   2014   &  59 & 56722--56972 & 0.0057 & $0.9546\pm0.0007$  \\
	   2015   &  73 & 57028--57335 & 0.0062 & $0.9549\pm0.0007$  \\
	   2016   &  26 & 57385--57451 & 0.0055 & $0.9485\pm0.0011$  \\
	\enddata
\end{deluxetable}

\section{\HST\ and \Spitzer\ Light Curve Fits}
\label{sec:fits}
We fit a two-component model to the light curves. One component models the astrophysical signal (the planet's thermal phase variation and transit), and the other component models the systematic noise introduced by time-dependent changes in instrument performance. For each light curve, we fit the physical and systematic components simultaneously, such that the total observed flux as a function of time is given by $F(t) = F_\mathrm{physical}(t) \times F_\mathrm{sys}(t)$. For the \HST\ data, where we observed two phase curves and two additional eclipses, we constrain the physical parameters to be the same for all visits, but allow some of the systematics parameters to vary (for more details see \S\,\ref{sec:hst_sys}). We fit the WFC3 band-integrated ``white" light curve, as well as spectroscopic light curves created from 10 wavelength bins uniformly spaced at $0.05\,\mu$m intervals between $1.15$ and $1.65\,\mu$m.

\subsection{Astrophysical Signal}
We assume the measured astrophysical signal $F_\mathrm{physical}$ has the following form:
\begin{equation}
	F_\mathrm{physical}(\lambda, t) =  T(\lambda, t) + c(\lambda, t) \times F_p/F_s(\lambda, t)
\end{equation}
where $\lambda$ is wavelength, $T(\lambda, t)$ is the transit model (the fraction of the stellar disk that is visible at time $t$), $F_p/F_s(\lambda, t)$ is the disk-integrated planet-to-star flux, and $c$ is a correction factor for companion star dilution and the planet's tidal distortion. 

We calculated the transit model $T(t)$ with the \texttt{batman} package \citep{kreidberg15a}. Many of the physical parameters are tightly constrained by \cite{southworth15}, so we fixed the orbital period, time of inferior conjunction, orbital inclination, and ratio of semi-major axis to stellar radius to the previously published values ($P = 0.925545613$ day, $t_0 = 2456836.2964455\,\mathrm{BJD_{TDB}}$, $i = 87.3^\circ$, and $a/R_s = 2.999$). As a test, we fit for these parameters with the \Spitzer\ Channel 2 light curve, which has the best phase coverage and least systematic noise of the three data sets. We found that the transit parameters are consistent with the \cite{southworth15} results, so we proceeded with the remainder of the analysis holding those parameters fixed.  The free parameters for the transit model were a wavelength-dependent transit depth $r_p(\lambda)$ and linear limb darkening parameter $u(\lambda)$. More complex limb darkening laws with additional free parameters were not merited according to the Bayesian Information Criterion (BIC). We initialized our MCMC chains on the least squares best fit parameters.

We modeled the planet-to-star flux $F_p/F_s$ in two different ways. First, we fit a sinusoid with a period equal to the planet's orbital period. The free parameters were the sine curve amplitude and phase offset. For the second approach, we used the \texttt{spiderman} package \citep{louden17} to model $F_p/F_s$. This package allows users to input a climate map (temperature or brightness as a function of latitude and longitude), and generate the corresponding flux ratio for an observation at time $t$.  In our fit, we calculated the stellar flux with a NextGen model \citep{allard12} interpolated to an effective temperature of $6110\,\mathrm{K}$ \citep{gillon14}, solar metallicity, and log\,$g$ of 4.2 (in cgs units).  For the planet flux, we tested three different maps: a two-temperature map, with a uniform dayside temperature $T_d$ and a uniform nightside temperature $T_n$; a spherical harmonic maps of degree two (with four free parameters); and the physically-motivated kinematic model from \cite{zhang17}, which has just three free parameters (the nightside temperature $T_n$, the change in temperature from day-to-night side $\Delta_T$, and the ratio of radiative to advective timescales $\xi$).  In all cases, we assumed that the planet is tidally locked, such that each orbital revolution corresponds to one complete rotation on its spin axis. 

\begin{figure}
\includegraphics[width = 0.5\textwidth]{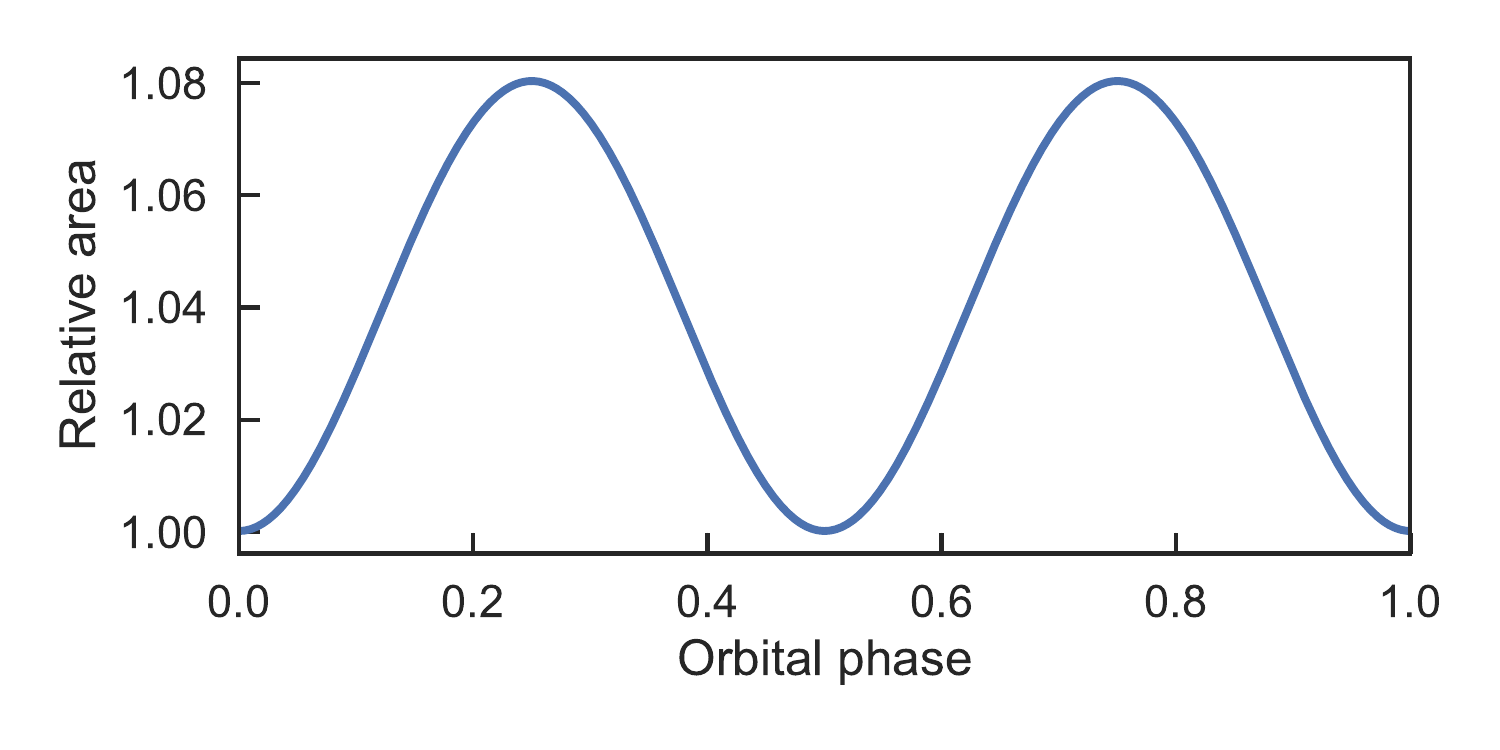}
\caption{Projected area of the planet as a function of orbital phase, normalized to unity at phase zero. The area variation was predicted analytically using the model from \cite{leconte11b}.}
\label{fig:ellipsoidal}
\end{figure}

We scaled the planet-to-star flux by a correction factor $c$ to account for dilution from the companion star and ellipsoidal variability due to the planet's tidal distortion. The correction factor took the form: 
\begin{equation}
	c(\lambda, t) = [1 + \alpha(\lambda)]A(t)
\end{equation}
where $\alpha(\lambda)$ is the additional fractional flux from the companion star and $A(t)$ is the sky-projected area of the planet. We estimated $\alpha(\lambda)$ based on the best fit spectral energy distribution from \cite{cartier17}. The companion star contribution ranges from $10\%$ at $1.1\,\mu$m to $16\%$ at $4.5\,\mu$m. The uncertainty on the companion star flux contribution to the total system flux is less than 1\%, which introduces negligible error in the estimated planet-to-star-flux compared to the photon noise.   We calculated $A(t)$ using the analytic formula from \cite{leconte11b}, equation B.9, which computes the projected area of a triaxial ellipsoid. We estimated the ellipsoid properties using Table B.3 of \cite{leconte11a}, assuming the planet radius is 1.5 $R_\mathrm{Jup}$ and age is 5 Gyr. The predicted ellipsoidal variability is shown in Figure\,\ref{fig:ellipsoidal}. At quadrature, the projected area is $8\%$ larger than at phase zero (mid-transit). Using the analytic expression from \cite{loeb03}, we estimated the effect of Doppler beaming and found that it contributes less than 10 ppm to the measured flux. 

\begin{figure}
    \includegraphics[width = 0.5\textwidth, trim={0cm 1cm 0cm 3cm},clip]{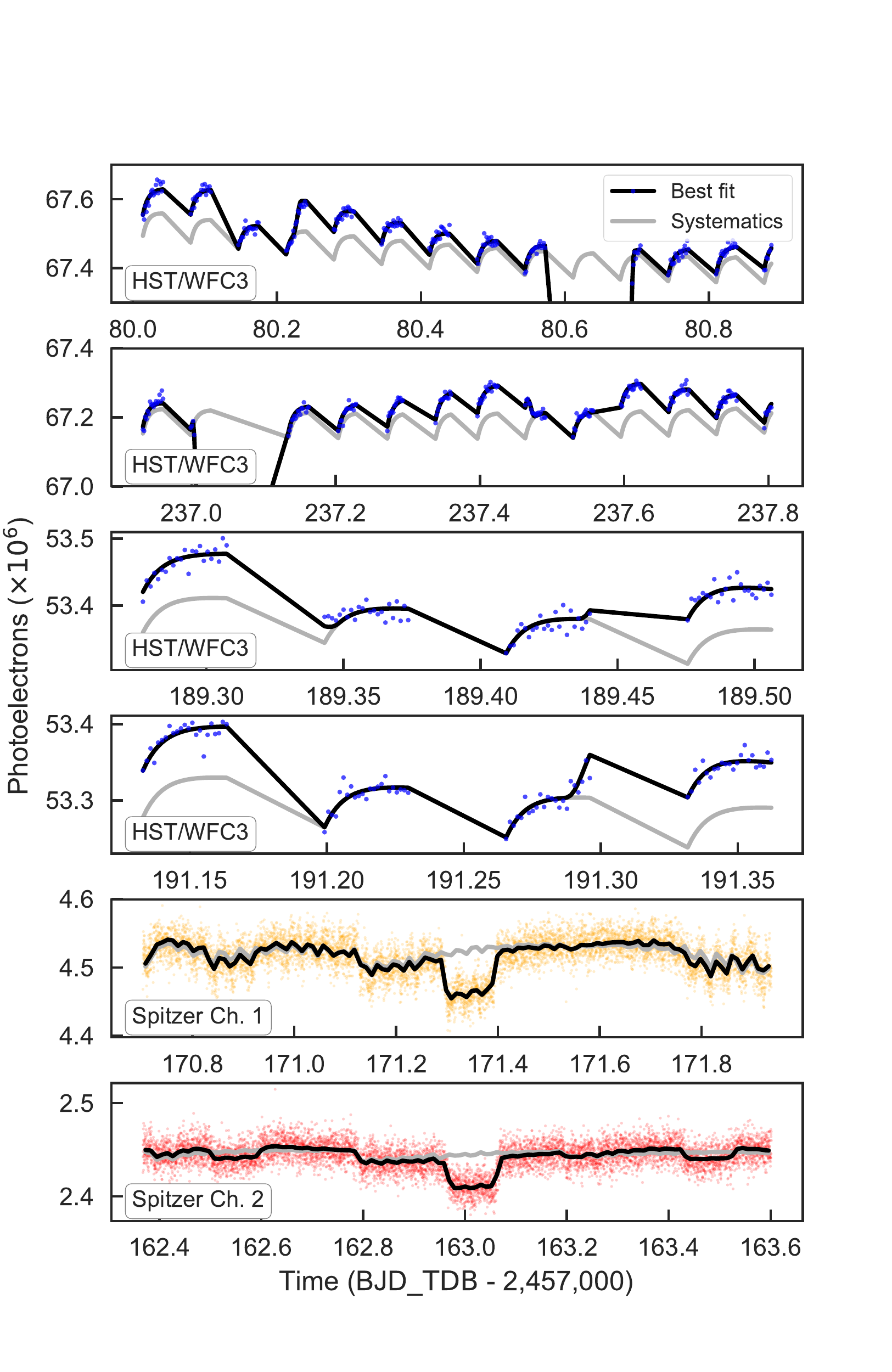}
\caption{Raw light curves (points) for WASP-103b observed with \HST/WFC3 (top four panels) and \Spitzer/IRAC (bottom two panels). The black lines show the best fit models, which include the astrophysical signal and instrument systematics. The gray lines indicate the contribution from the instrument systematics alone (which would be observed for a source with constant brightness and no planet). For visual clarity, we corrected the \HST\ data for the upstream-downstream effect and zoomed in on the phase variation, so the transits are not displayed in the panel.}
\label{fig:systematics}
\end{figure}

\subsection{Systematics}
Both the \HST\ and \Spitzer\ phase curves have systematic noise caused by variations in the sensitivity of the instrument over time. For the \HST/WFC3 data, the dominant systematic is an orbit-long exponential trend due to charge traps filling up over successive exposures \citep{long15, zhou17}. For \Spitzer\, the primary source of noise is the intrapixel sensitivity effect. The detector's pixels do not have uniform sensitivity, so slight changes in telescope pointing cause the recorded flux to vary. In Figure\,\ref{fig:systematics}, we show the raw light curves before systematic noise was removed. The systematics have comparable amplitude to the thermal phase variation signal, so they must be carefully corrected to recover the underlying planet-to-star flux. 

\subsubsection{\HST\ Systematics}
\label{sec:hst_sys}
We fit the WFC3 systematics using an analytic model of the form:
\begin{equation}
 F_\mathrm{sys}(t) = (c\,S(t) + v_1\,t_\mathrm{v} + v_2\,t_\mathrm{v}^2)(1 - \exp(-a\,t_\mathrm{orb} - b))
\end{equation}
where $t_\mathrm{v}$ is time elapsed since the first exposure in a visit and $t_\mathrm{orb}$ is time since the first exposure in an orbit. $S(t)$ is a scale factor equal to 1 for exposures with spatial scanning in the forward direction and $s$ for reverse scans, to account for the upstream-downstream effect \citep{mccullough12}. The orbit-long ramp parameters are consistent for all the visits, so we constrained $a$, $b$, and $s$ to have the same value for all visits in the final fit. The visit-long trends differ from visit to visit, so $c$, $v_1$, and $v_2$ were allowed to vary between visits. We fixed $v_2$ to zero for the two secondary eclipse observations from Program 13360, since the visit-long trend for shorter observations is fit well by a linear slope.

Some segments of the data exhibit stronger systematics than others, so we exclude these data in our final analysis. We drop the first orbit from every visit and the first exposure from every orbit \citep[following common practice; see e.g.][]{kreidberg14a}.  We also discard exposures from the last half of orbit 15 from the phase curve observations, which were taken in staring mode to enable a gyro bias update. Since we observed two phase curves, we have complete orbital phase coverage of the planet despite discarding some data.

\subsubsection{\Spitzer\ Systematics}
Warm Spitzer's primary systematic is intrapixel sensitivity variation, where the photometry depends on the precise location of the stellar center within its pixel.  We fit this systematic using the Bilinearly-Interpolated Subpixel Sensitivity (BLISS) mapping technique \citep{stevenson12}.  BLISS provides a flexible, non-analytic means to effectively weight the target flux by the spatial sensitivity variations within a pixel, while simultaneously fitting for other systematics and the physical parameters of the system. As demonstrated by \citet{Ingalls2016}, the POET pipeline with BLISS mapping can accurately model simulated \Spitzer\ light curves with known physical parameters and produce reliable results.

The BLISS sensitivity map is determined by bilinear interpolation over a grid of knots centered on the stellar flux. Each knot's sensitivity is calculated from the residuals to the light curve fit: the higher the flux values for data points near a given knot, the higher the detector sensitivity is at that position.  To avoid overfitting, we chose the grid scale such that bilinear interpolation performed better than nearest neighbor interpolation. For the $3.6\,\mu$m data, the grid scale was 0.008 pixel (0.0098 arcsec) in both $x$ and $y$. For $4.5\,\mu$m, the scale was 0.022 pixel (0.027 arcsec).  In addition to the intrapixel sensitivity variation, we fit the data for a linear trend in time. We tested a quadratic trend but did not find significant evidence for the additional model complexity based on the BIC.  

\subsection{Best Fits and Uncertainties}
To determine the best fits, we performed a least-squares $\chi^2$ minimization for each wavelength and model. For a subset of these cases where we wish to calculate 68\% confidence intervals, we also performed a Markov chain Monte Carlo (MCMC) analysis to estimate parameter uncertainties. These include the transit fits and the sine curve fit to the broadband phase curves. We used \texttt{emcee} \citep{foremanmackey13} to fit the \HST/WFC3 light curves and differential evolution Monte Carlo for the \Spitzer\ fits \citep{braak06}. We ran the MCMC until convergence according to the Gelman-Rubin statistic. We initialized the MCMC chain on the best fit parameters and discarded the first 10\% of the chain as burn-in. MCMC techniques only produce robust uncertainties when the noise is normally distributed and white, so to account for correlated noise in the $3.6\,\mu$m light curve (described in \S\,\ref{sec:fitquality}) we fit the wavelet model from \cite{carter09} simultaneously with the other model parameters. We used a Haar wavelet and let the power spectral density of the red noise vary, following \cite{diamond-lowe14}. In our final fit, the noise power spectrum $1/f^\gamma$ had $\gamma = 1.1 \pm 0.1$ (implying an equal amount of white noise and correlated noise).  
\subsection{Goodness of Fit}
\label{sec:fitquality}
We performed several tests of the quality of the light curve fits.  First we
predicted the level of scatter in the light curves based on photon noise alone,
then compared this value to the root-mean-square (rms) of the fit residuals.
For the spherical harmonics fit to the phase variation, the Spitzer 4.5 $\mu$m
light curve rms reaches the expected photon noise limit (637 versus 640 ppm).
The 3.6 $\mu$m light curve has significantly larger rms (767 versus 470 ppm),
due to time-correlated red noise (discussed below). The expected photon-limited
rms for the WFC3 spectrosopic light curves ranges from 430 - 530 ppm, and the
measured rms was typically within 5\% of expectations for all spectroscopic
channels.  For the WFC3 band-integrated white light curve, the rms was slightly
larger than predicted (172 versus 122 ppm). There are a number of possible
origins for this discrepancy, including imperfect background subtraction,
variation in the position of the spatial scan on the detector, and loss of flux
outside the extraction window. In addition, the amplitude of the phase
variation increases by 50\% over the WFC3 wavelength range, which leads to a
small increase in the noise in the white light curve. To make an order of
magnitude estimate for the amplitude of this effect, we calculated the standard
deviation of the secondary eclipse depths in all wavelength channels. It is
$\sim100$ ppm, which is comparable to the additional scatter we observed in the
white light curve. 



\begin{figure}
\includegraphics[width = 0.5\textwidth]{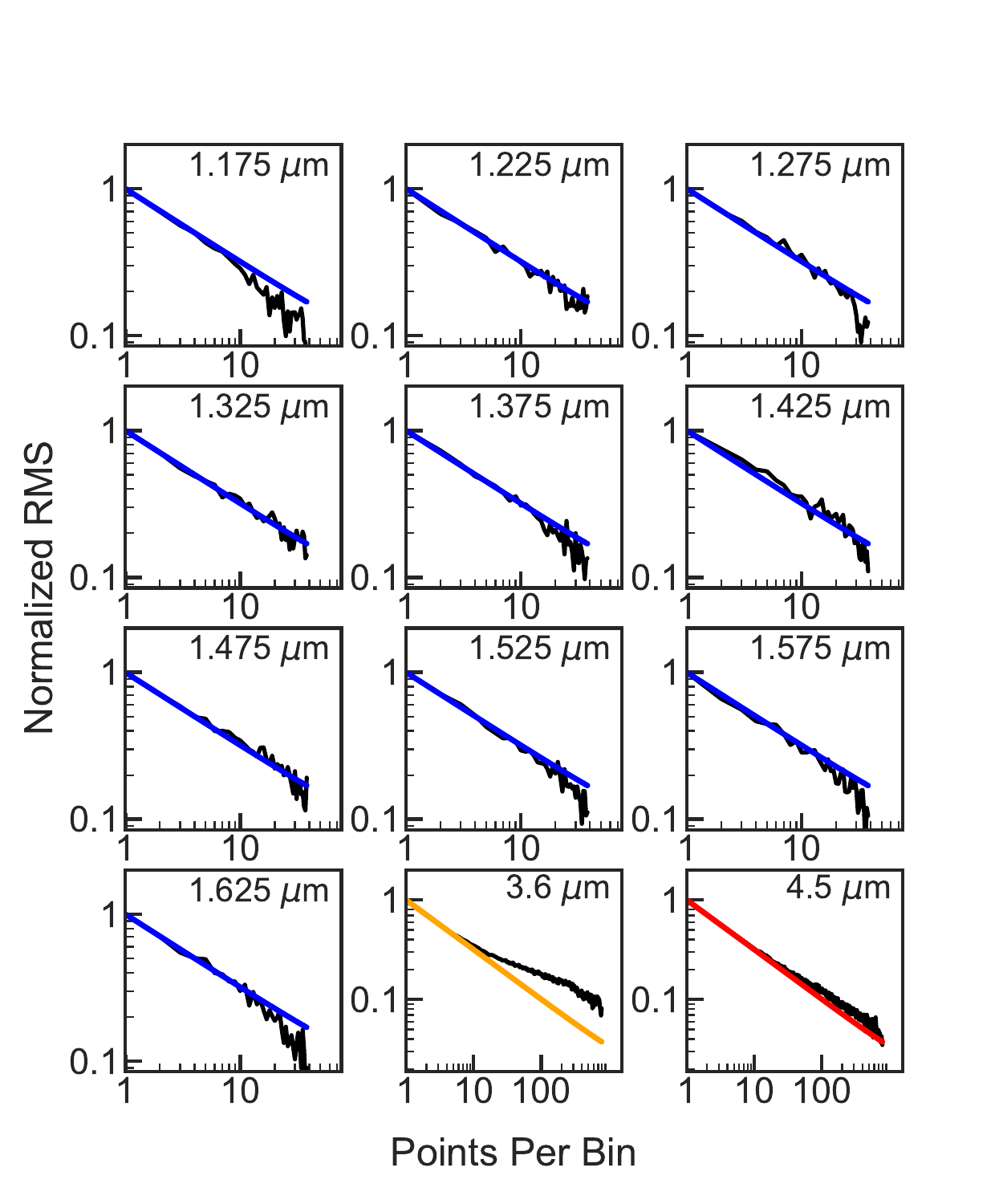}
\caption{Root mean square (rms) variability in the light curves as a function of bin size (black lines) compared to the expected rms from photon noise (colored lines). The central wavelength of the light curve is indicated in the upper right corner of each panel. With the exception of the \Spitzer\ 3.6 $\mu$m channel, the rms for the light curves bins down in agreement with predictions from the photon noise.}
\label{fig:rms}
\end{figure}

\begin{figure*}
\includegraphics[width = 1.0\textwidth]{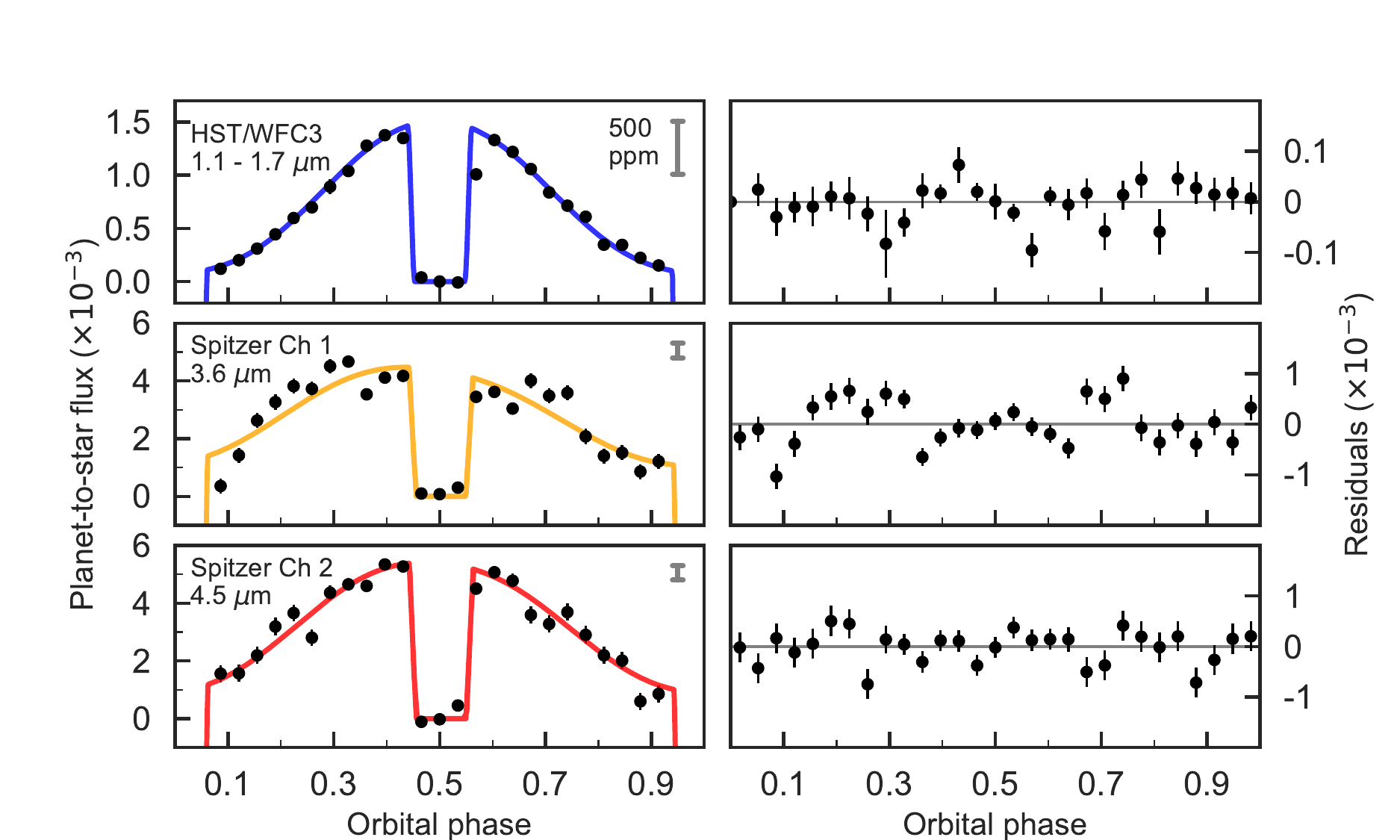}
\caption{WASP-103b phase curve observations from \HST/WFC3 (top) and
\Spitzer/IRAC (middle and bottom). For clarity, the data are phase-folded on the
planet's orbital period and binned in 30 uniformly spaced bins between 0 and 1
(corresponding to 0.8 hours). The left column shows the phase curves with systematic noise removed (black points) compared to the best fit spherical harmonics model (colored lines). The error bars denote $1\,\sigma$ uncertainties (in some cases, the errors are smaller than the data points).  We include the transits in the fit, but they are not displayed in this figure. The right-hand column shows the binned residuals for the best-fit light curve. The gray error bars in the upper right of the left panels correspond to 500 ppm, to illustrate the changing y-axis scale.}
\label{fig:phasecurves}
\end{figure*}

In addition to calculating the fit rms compared to the photon noise, we also tested for the presence of red noise based on whether the rms decreases as expected when the light curve in binned in time.  If the noise is white (uncorrelated in time), the residuals are expected to decrease by a factor of $\sqrt{N}$, where $N$ is the number of points in a bin. Figure\,\ref{fig:rms} shows the binned residuals compared to expectations for white noise. The \HST/WFC3 and \Spitzer\ Channel 2 light curves agree well with expectations, whereas \Spitzer\ Channel 1 shows higher noise than expected as bin size increases. This test confirms the presence of time-correlated noise in the Channel 1 light curve that can be seen by eye in the residuals in Figure\,\ref{fig:phasecurves}. Both \Spitzer\ channels use the same detector, but Channel 1 data are more susceptible to time-correlated noise because the the point spread function is more undersampled at shorter wavelengths, making intrapixel sensitivity variations more pronounced.

\section{Results}
\label{sec:results}
The fitted light curves are shown in Figure\,\ref{fig:phasecurves}. This figure shows results from the spherical harmonics model for the thermal phase variation and has instrument systematics removed.  Broadly speaking, the phase curves show large dayside planet-to-star flux values, ranging from $0.151\pm0.015\%$, $0.446\pm0.38\%$, and $0.569\pm0.014\%$ in the WFC3 white light curve, and \Spitzer\ $3.6$, and $4.5\mu$m bandpasses, respectively.  The planet flux changes significantly with orbital phase in all three of the data sets, suggesting a strong gradient from dayside to nightside temperature, and peak brightness occurs near phase 0.5. In this section, we quantitatively characterize the phase curve shape, split the data into phase resolved spectra, evaluate different temperature maps, compare with previous observations of the dayside thermal emission spectrum, and report the transmission spectrum.

\subsection{Phase Curve Amplitudes and Hotspot Offsets}
The shape of a phase curve can be summarized with two parameters: the amplitude of thermal phase variation (minimum to maximum brightness, divided by the secondary eclipse depth) and the location of peak brightness (typically called a ``hotspot offset" and measured in degrees eastward of the substellar point).  Table\,\ref{table:amps_offsets} lists the estimated amplitudes and hotspot offsets (median and $1\,\sigma$ credible interval) for the band-integrated WFC3 phase curve and both \Spitzer\ channels. The estimates are from the sine curve model for the thermal phase variation. The advantage of using this model (even though it does not provide strictly the best fit), is that it fits for the amplitude and offset directly as free parameters. 


For all three phase curves, the hotspot offset is consistent with zero degrees, which could indicate a small ratio of radiative to advective timescales (the incident flux is reradiated to space faster than it is advected around to the nightside). Fast radiative timescales are predicted at high temperatures, and small hotspot offsets are also observed for other very hot Jupiters \citep{perez13, komacek16, komacek17}.  The measured offsets are inconsistent with the trend reported in \cite{zhang18}, which predicts the hotspot offset increases with planet temperature for irradiation temperatures greater than 3410 K. The \cite{zhang18} model predicts an eastward hotspot offset of $4.5^\circ$ for WASP-103b, which is significantly larger than observed, hinting at diversity in the circulation patterns of the hottest planets.  The phase curve amplitudes are large (near 0.8 - 0.9), as expected for an atmosphere with inefficient heat redistribution. In \S\,\ref{sec:gcm} we compare these results to expectations from 3D GCMs.


\begin{deluxetable}{llLL}
	\tablecolumns{4}
	\tablewidth{0pt}
	\tablecaption{Phase Curve Properties \label{table:amps_offsets}}
	\tablehead{
		\colhead{Bandpass} & \colhead{Source} & \colhead{Amplitude} & \colhead{Offset}\\
	\colhead{\,} & \colhead{\,} & \colhead{\,} & \colhead{(Degrees)}}
		\startdata
		WFC3 & data & 0.91 \pm 0.02 & -0.3 \pm 0.1 \\
		\, & nominal GCM & 0.89 & 15.32 \\
		\, & [M/H] = 0.5 GCM & 0.84 & 19.64 \\
		\, & $\tau_\mathrm{drag4}$ GCM & 0.97 & 2.34 \\
		\, & $\tau_\mathrm{drag3}$ GCM & 0.99 & 0.18 \\
		Spitzer 3.6 $\mu$m & data & 0.86 \pm 0.13 & 2.0 \pm 0.7 \\
		\, & nominal GCM & 0.78 & 9.19 \\
		\, & [M/H] = 0.5 GCM & 0.72 & 12.79 \\
		\, & $\tau_\mathrm{drag4}$ GCM & 0.86 & 0.90 \\
		\, & $\tau_\mathrm{drag3}$ GCM & 0.97 & 0.18 \\
		Spitzer 4.5 $\mu$m & data & 0.83 \pm 0.05 & 1.0 \pm 0.4 \\
		\, & nominal GCM & 0.79 & 8.11 \\
		\, & [M/H] = 0.5 GCM & 0.73 & 11.35 \\
		\, & $\tau_\mathrm{drag4}$ GCM & 0.85 & 0.90 \\
		\, & $\tau_\mathrm{drag3}$ GCM & 0.93 & 0.18 \\
		\enddata
	\end{deluxetable}

\subsection{Phase-Resolved Spectra}
We used the best-fit phase curves (with systematics removed) to generate phase-resolved emission spectra.  Since the \texttt{spiderman} thermal phase variation models fit the temperature of the planet rather than the eclipse depth directly, we estimated the dayside emission spectrum as follows. We used \texttt{spiderman}'s eclipse\_depth method to calculate the average planet-to-star flux for the best-fit model during secondary eclipse.  To estimate uncertainties, we took the standard deviation of the residuals of the in-eclipse data points, then added this value in quadrature to the standard deviation of the residuals of the out-of-eclipse data.  This quadrature sum accounts for the uncertainty in the baseline flux. To account for red noise in the \Spitzer\ $3.6\,\mu$m light curve, we use the approach of \cite{pont06} to determine the red noise contribution on the timescale of the eclipse. We add the estimated red noise in quadrature, which increases the uncertainty on planet-to-star flux by a factor of $2.5$.

\begin{figure*}
\includegraphics[width = 1.0\textwidth]{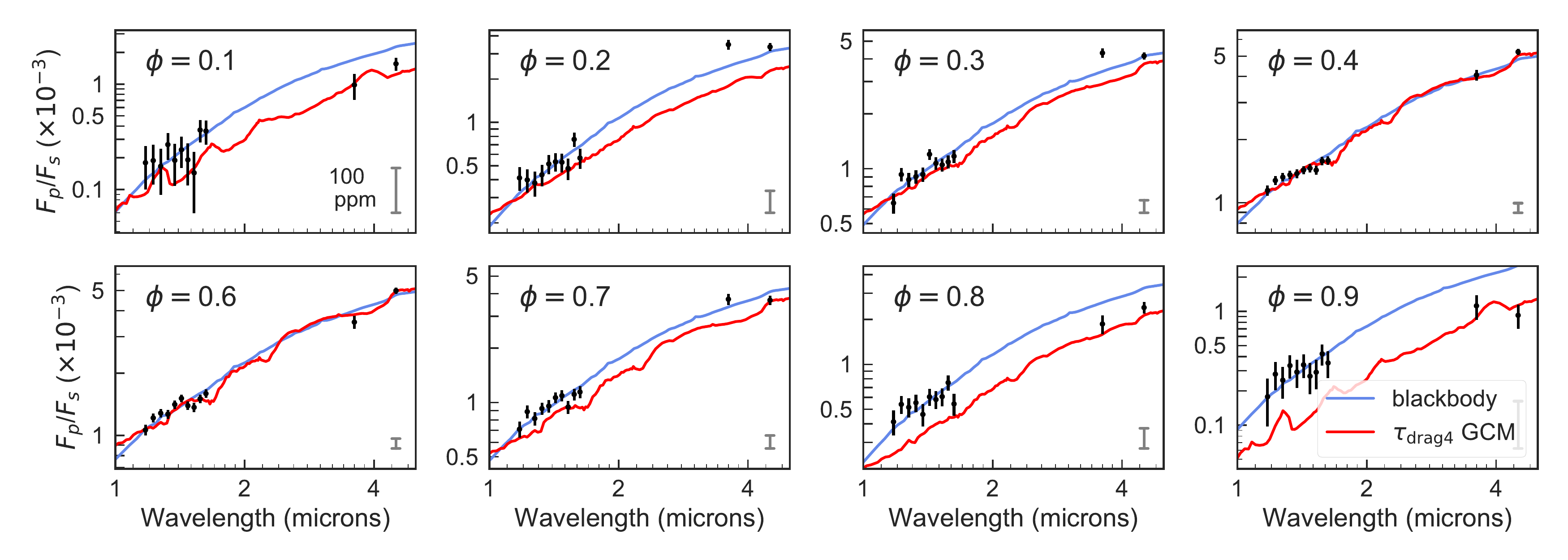}
\caption{Phase-resolved planet-to-star flux ratios (points) compared to the best fit blackbody (blue line) and the GCM with $\tau_\mathrm{drag4}$. Phases $\phi=0.5$ and $0.0$ correspond to times of secondary eclipse and transit, when the substellar and antistellar points are respectively aimed at Earth. The blackbody model is fit to the \HST/WFC3 data only. We also show a 100 ppm error bar in the lower right of each panel to emphasize the changing limits on the y-axes.}
\label{fig:spectra}
\end{figure*}

\begin{figure*}
\includegraphics[width = 1.0\textwidth]{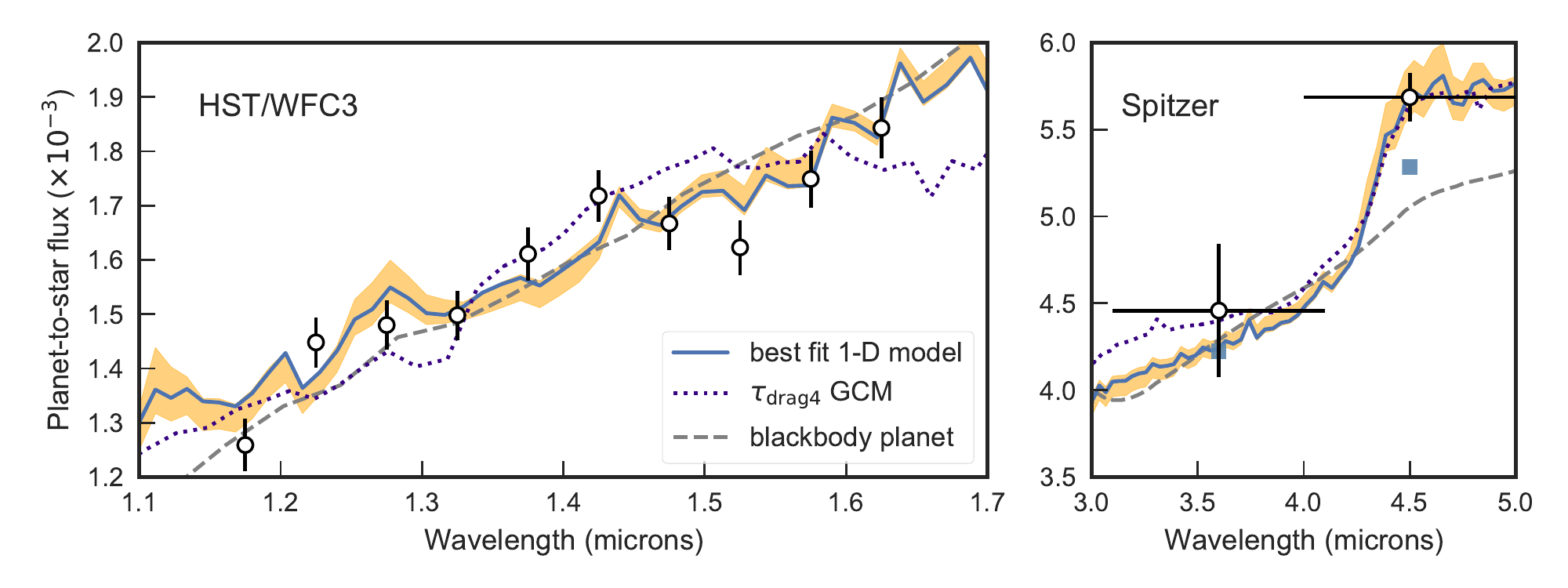}
\caption{Dayside planet-to-star flux (points) compared to a blackbody model for the planet (dashed gray line) and the best fit 1D model (blue line, with $1\,\sigma$ uncertainty shaded in orange) for the \HST/WFC3 data (left) and Spitzer (right). The 1D model was fit to all the data simultaneously, but the blackbody was fit to the WFC3 data only. The best fit 1D model binned at the resolution of the \Spitzer\ data is indicated by blue squares (right).} 
\label{fig:dayside}
\end{figure*}

For the other orbital phases, we binned the light curve (with systematics
removed) in eight intervals of about 0.1 in orbital phase (2.2 hours), with endpoints at
phases $0.06, 0.15, 0.25, 0.35, 0.44$ and $0.56, 0.65, 0.75, 0.85, 0.94$. These
endpoints were chosen to ensure that there is no contribution from in-transit or
in-eclipse data.  In each phase bin, we estimated the planet-to-star flux from
the mean value of the data points in the bin. To estimate the uncertainty, we
took the standard deviation of the points in the bin and added it in quadrature
to the standard deviation of the data points during secondary eclipse (phase
$0.46-0.54$), to account for the uncertainty in baseline stellar flux.  For the
$3.6\mu$m data, we also add red noise on the timescale of a phase bin, following
\cite{pont06}.  The phase-resolved emission spectra are shown in
Figure\,\ref{fig:spectra} and listed in Table\,\ref{table:spectra}. We show the
dayside spectrum in Figure\,\ref{fig:dayside}.

\begin{figure*}
\includegraphics[width = 1.0\textwidth, trim={1.5cm 0 0.5cm 0},clip]{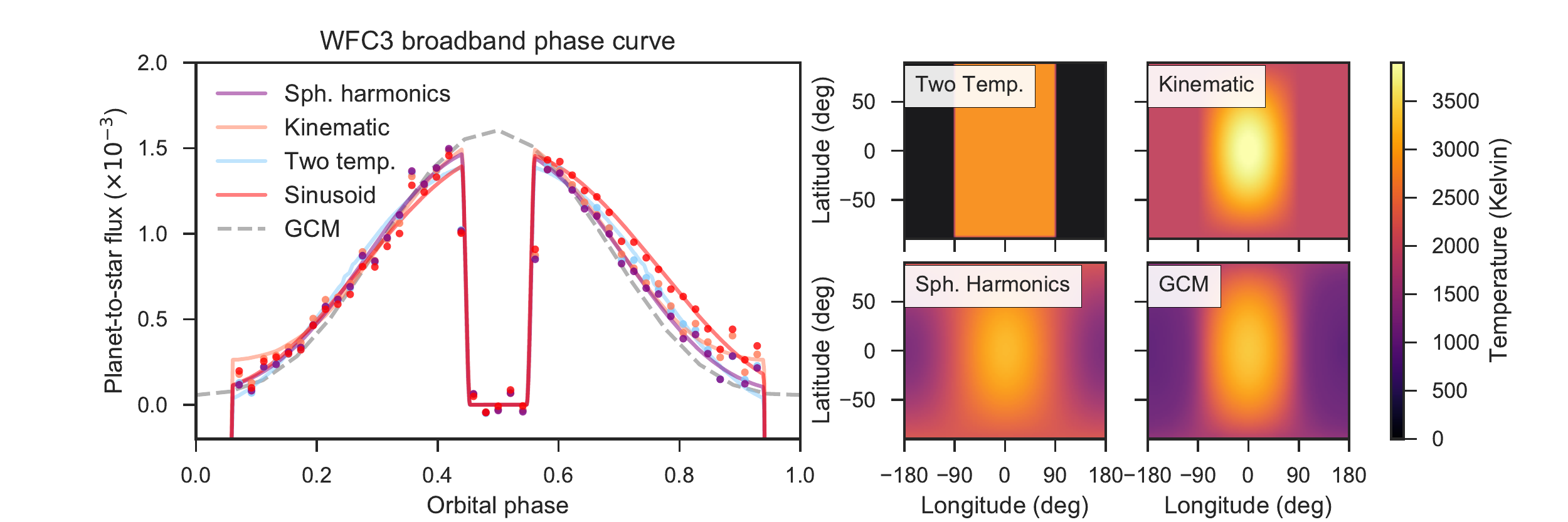}
\caption{\textbf{Left:} Fits to the broadband WFC3 phase curve compared to a GCM. The colored lines correspond to different temperature maps fit to the data, and the dashed gray line is from the $\tau_\mathrm{drag4}$ GCM. We also show the measured planet-to-star flux for each map (points), which is model-dependent due to slight degeneracies with the instrument systematic model.  \textbf{Right:} Temperature maps from the best fit models and the GCM at a pressure of 0.1 bar.}
\label{fig:model_comparison}
\end{figure*}

To test that the phase-resolved planet-to-star flux values are robust to different approaches for fitting the phase curves, we compared the estimated planet-to-star flux for all four of the thermal phase variation models (sinusoid, kinematic, spherical harmonics, and two temperature).  Since the systematic noise is not strongly correlated with the astrophysical signal, the systematics-divided data are nearly identical for all the models.  This point is illustrated in Figure\,\ref{fig:model_comparison} for the broadband WFC3 light curve.  We found that the choice of model generally does not significantly change the estimated planet-to-star flux ratios.  The estimates agree to better than one sigma for 90\% of phase bins for the spherical harmonics, two temperature, and physical models. For the WFC3 data, the sinusoid is higher than the other models by an average of $1.5\,\sigma$ for phases $0.5 - 1$. This discrepancy may be due to the added flexibility in hotspot offset for the sinusoid model; other models do not allow for westward hotspot offsets.

\begin{deluxetable*}{lllllllllll}
	\tablecolumns{11}
	\tablewidth{0pt}
	\tablecaption{Phase-Resolved Emission Spectra \label{table:spectra}}
	\tablehead{
	\colhead{$\lambda$} & \colhead{Dilution} & \colhead{$\phi = 0.1$} & \colhead{$\phi = 0.2$} & \colhead{$\phi = 0.3$} & \colhead{$\phi = 0.4$} & \colhead{$\phi = 0.5$} & \colhead{$\phi = 0.6$} & \colhead{$\phi = 0.7$} & \colhead{$\phi = 0.8$} & \colhead{$\phi = 0.9$}}
		\startdata
		1.175 & 0.10 & $ 179 \pm 79 $ & $ 411 \pm 77 $ & $ 647 \pm 80 $ & $ 1143 \pm 65 $ & $ 1259 \pm 47 $ & $ 1063 \pm 64 $ & $ 710 \pm 73 $ & $ 412 \pm 78 $ & $ 177 \pm 79 $ \\ 
		1.225 & 0.11 & $ 188 \pm 76 $ & $ 398 \pm 74 $ & $ 928 \pm 77 $ & $ 1276 \pm 62 $ & $ 1448 \pm 46 $ & $ 1216 \pm 62 $ & $ 888 \pm 71 $ & $ 539 \pm 75 $ & $ 280 \pm 76 $ \\ 
		1.275 & 0.11 & $ 166 \pm 76 $ & $ 379 \pm 74 $ & $ 869 \pm 77 $ & $ 1323 \pm 62 $ & $ 1480 \pm 46 $ & $ 1282 \pm 62 $ & $ 814 \pm 71 $ & $ 515 \pm 75 $ & $ 247 \pm 76 $ \\ 
		1.325 & 0.11 & $ 266 \pm 75 $ & $ 432 \pm 73 $ & $ 904 \pm 76 $ & $ 1357 \pm 62 $ & $ 1498 \pm 45 $ & $ 1267 \pm 61 $ & $ 925 \pm 70 $ & $ 552 \pm 74 $ & $ 333 \pm 75 $ \\ 
		1.375 & 0.12 & $ 189 \pm 81 $ & $ 514 \pm 78 $ & $ 928 \pm 82 $ & $ 1376 \pm 66 $ & $ 1611 \pm 48 $ & $ 1411 \pm 65 $ & $ 954 \pm 75 $ & $ 461 \pm 79 $ & $ 292 \pm 81 $ \\ 
		1.425 & 0.13 & $ 238 \pm 79 $ & $ 532 \pm 76 $ & $ 1198 \pm 79 $ & $ 1431 \pm 64 $ & $ 1718 \pm 47 $ & $ 1511 \pm 64 $ & $ 1063 \pm 73 $ & $ 605 \pm 77 $ & $ 338 \pm 79 $ \\ 
		1.475 & 0.14 & $ 191 \pm 81 $ & $ 527 \pm 79 $ & $ 1068 \pm 82 $ & $ 1460 \pm 66 $ & $ 1667 \pm 48 $ & $ 1392 \pm 66 $ & $ 1090 \pm 75 $ & $ 580 \pm 80 $ & $ 268 \pm 81 $ \\ 
		1.525 & 0.14 & $ 143 \pm 84 $ & $ 478 \pm 81 $ & $ 1048 \pm 85 $ & $ 1429 \pm 69 $ & $ 1623 \pm 50 $ & $ 1367 \pm 68 $ & $ 943 \pm 77 $ & $ 607 \pm 82 $ & $ 291 \pm 84 $ \\ 
		1.575 & 0.15 & $ 367 \pm 88 $ & $ 761 \pm 85 $ & $ 1088 \pm 89 $ & $ 1581 \pm 72 $ & $ 1749 \pm 52 $ & $ 1503 \pm 71 $ & $ 1107 \pm 81 $ & $ 754 \pm 86 $ & $ 422 \pm 88 $ \\ 
		1.625 & 0.16 & $ 359 \pm 93 $ & $ 565 \pm 90 $ & $ 1169 \pm 94 $ & $ 1590 \pm 76 $ & $ 1843 \pm 56 $ & $ 1593 \pm 75 $ & $ 1142 \pm 86 $ & $ 542 \pm 91 $ & $ 351 \pm 93 $ \\ 
		3.6 & 0.17 & $ 982 \pm 271 $ & $ 3474 \pm 268 $ & $ 4309 \pm 255 $ & $ 4060 \pm 248 $ & $ 4458 \pm 383 $ & $ 3524 \pm 249 $ & $ 3725 \pm 267 $ & $ 1865 \pm 269 $ & $ 1116 \pm 272 $ \\ 
		4.5 & 0.16 & $ 1560 \pm 220 $ & $ 3347 \pm 213 $ & $ 4150 \pm 189 $ & $ 5240 \pm 178 $ & $ 5686 \pm 138 $ & $ 4995 \pm 181 $ & $ 3677 \pm 212 $ & $ 2403 \pm 213 $ & $ 921 \pm 219 $ \\ 
		\enddata
		\tablecomments{The planet-to-star flux in each phase bin $\phi$ is in units of ppm.}
	\end{deluxetable*}

\begin{deluxetable}{lllllll}
	\tablecolumns{7}
	\tablewidth{0pt}
	\tablecaption{Model Comparison \label{table:models}}
	\tablehead{
	\colhead{Data} & \colhead{Model} & \colhead{$T_\mathrm{min}$} & \colhead{$T_\mathrm{max}$} & \colhead{$\overline{T}_\mathrm{night}$} & \colhead{$\overline{T}_\mathrm{day}$} & \colhead{$\Delta_\mathrm{BIC}$}}
		\startdata
		WFC3 & Sph. Harmonics & 1227 & 3237 & 1822 & 2636 & 0 \\
		\, & Kinematic & 1977 & 3953 & 1977 & 2769 & 14 \\
		\, & Two Temp. & 0 & 2879 & 0 & 2879 & 42 \\
		\, & Sinusoid & -- & -- & -- & -- & 17 \\
		Ch 1 & Sph. Harmonics & 1269 & 3391 & 1912 & 2741 & 0 \\
		\, & Kinematic & 1932 & 3630 & 1975 & 2614 & 34 \\
		\, & Two Temp. & 1418 & 2990 & 1418 & 2990 & 11 \\
		\, & Sinusoid & -- & -- & -- & -- & 25 \\
		Ch 2 & Sph. Harmonics & 888 & 3714 & 1729 & 2864 & 2 \\
		\, & Kinematic & 1614 & 3931 & 1621 & 2544 & 15 \\
		\, & Two Temp. & 1344 & 3241 & 1344 & 3241 & 0 \\
		\, & Sinusoid & -- & -- & -- & -- & 22 \\
		\enddata
		\vspace{-0.8cm}
		\tablecomments{comments}
	\end{deluxetable}

\subsection{Comparison with Previous Eclipse Observations}
We compared our results to the dayside emission spectrum reported by \cite{cartier17}, which is based on a Gaussian process analysis of two secondary eclipses from \HST/WFC3. The shape of their spectrum is consistent with what we find, but their eclipse depths are 125 ppm smaller on average (a difference of about 10\%). A likely explanation for this difference is that the \cite{cartier17} analysis does not include the planet's thermal phase variation, so that the Gaussian process models it as an instrument systematic.  If the phase variation is absorbed into the systematic model, the measured eclipse depths would be biased low.  By visual inspection of Figure\,\ref{fig:model_comparison}, we estimate the amplitude of this effect is $\sim100$ ppm, which is comparable to the offset between the two analyses. Our estimated uncertainties are a factor of four smaller than those reported in \cite{cartier17}, which is consistent with photon-limited expectations (our data set includes two additional eclipses, a factor of five longer out-of-eclipse baseline, and has 60\% larger wavelength bins).


We also compared our dayside spectrum to the $z'$ and $K_\mathrm{S}$-band secondary eclipse depths reported in \cite{delrez18}. The $z'$ (0.9 $\mu$m) eclipse is $1.0\,\sigma$ lower than our best fit blackbody spectrum (described in \S\,\ref{sec:composition}), and the $K_\mathrm{S}$ (2.1 $\mu$m) measurement is higher than the model by $2.5\,\sigma$. Since these results are consistent with (but less precise than) the WFC3 data, we do not include them in our analysis of the atmospheric composition, but we encourage additional measurements in the $K_\mathrm{S}$ band to confidently determine whether an emission feature is present at those wavelengths.

\subsection{Transmission Spectrum}
\begin{figure}
\includegraphics[width = 0.5\textwidth]{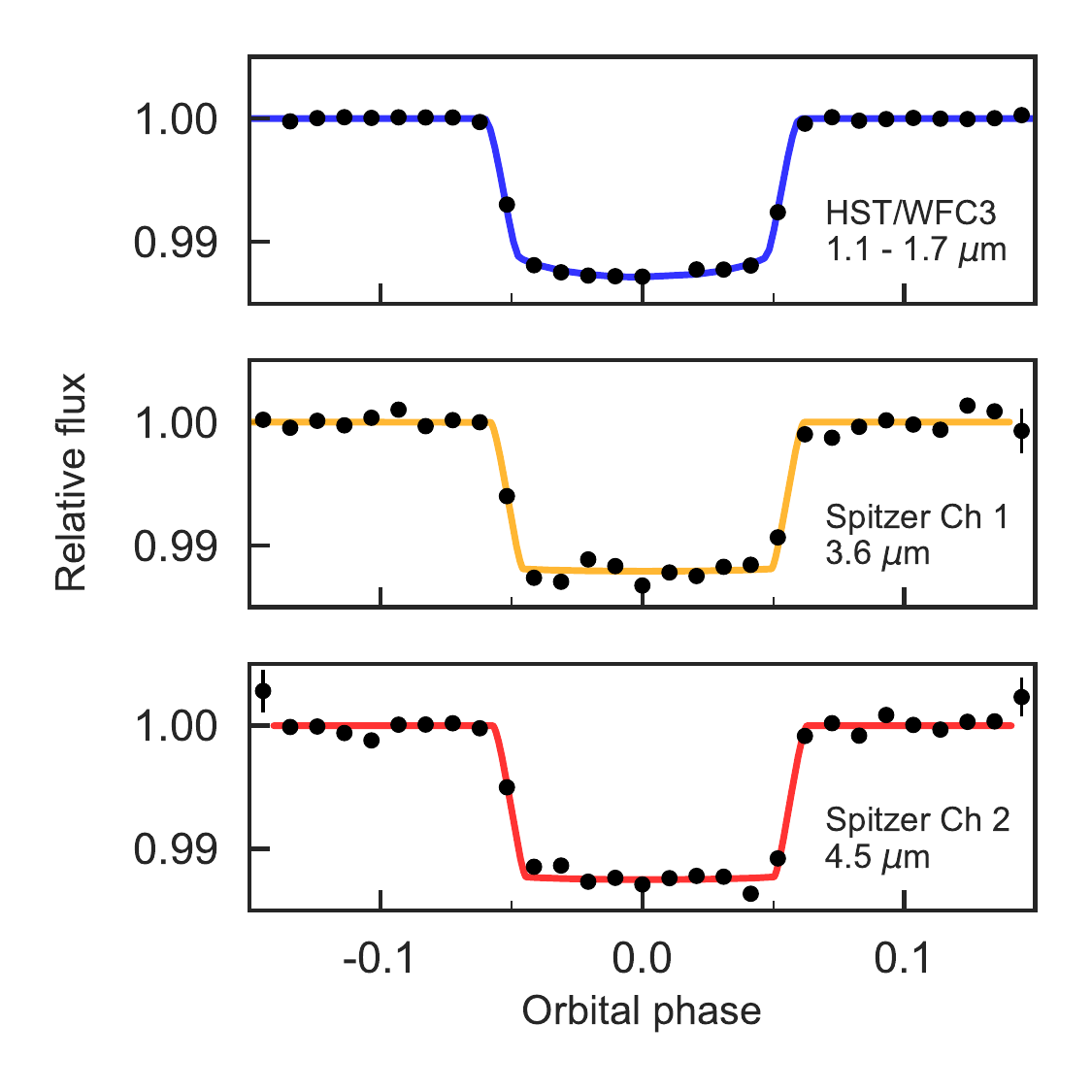}
\caption{Best fit transit light curves for WASP-103b (lines), compared to binned data (black points). From top-to-bottom, we show the band-integrated WFC3 light curve, \Spitzer\ 3.6, and $4.5\,\mu$m.}
\label{fig:transit_light_curves}
\end{figure}

Each phase curve observation includes a transit of WASP-103b. To measure the
wavelength-dependent transit depths (the transmission spectrum), we select a
subset of each phase curve. The subset includes the transit and additional
baseline on either side, such that the total light curve has twice the duration
of the transit. Over this short duration, there is negligible curvature in the
light curve due to the planet's thermal phase variation. We fit the data with a
transit model, which has free parameters for the planet-to-star radius and a
linear limb darkening parameter. The orbital parameters (inclination, $a/R_s$,
period, and time of central transit) were fixed at previously published values
listed in \S\,\ref{sec:fits}. We fit for the instrument systematics using the
same model as for the phase curve fits, except that we modeled the visit-long
systematic as a linear trend in time (which is sufficient for the shorter
duration). The advantage of fitting the transits separately from the full phase
curves is that the resultant transit depths are not dependent on how the phase
variation is modeled. The transit light curve fits are shown in
Figure\,\ref{fig:transit_light_curves}.

\begin{figure}
\includegraphics[width = 0.5\textwidth]{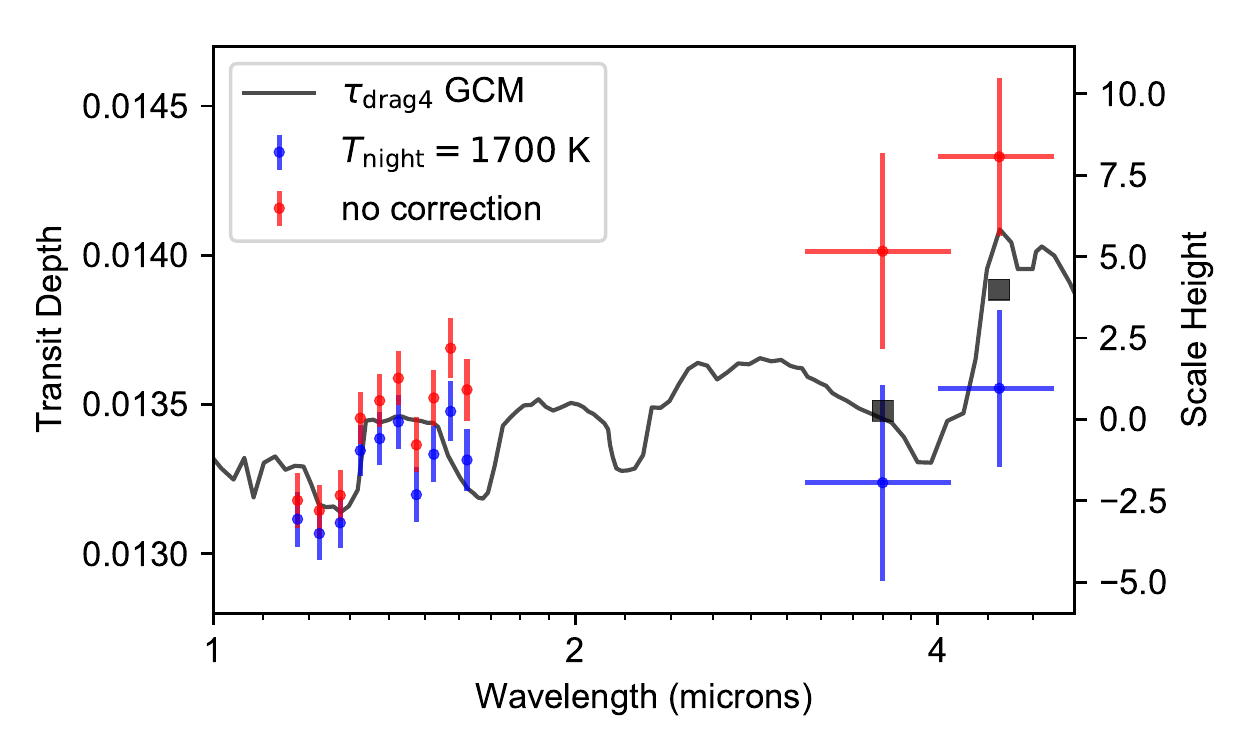}
\caption{The transmission spectrum of WASP-103b, corrected for planet nightside emission at 1700 K (blue points) and uncorrected (red points). The dark gray line is the model transmission spectrum from the $\tau_\mathrm{drag4}$ GCM, with squares indicating the model binned over the \Spitzer\ bandpasses. Atmospheric scale height $H$ is shown on the right y-axis, where $H = 5.5\times10^6$ m (assuming a mean molecular weight of 2.3 atomic mass units, surface gravity $g = 15.9$ m/s$^2$, and a temperature $T = 2410$ K).}
\label{fig:tspectrum}
\end{figure}

We show the measured transmission spectrum in Figure\,\ref{fig:tspectrum}.  The spectrum is biased by flux contamination from the planet's nightside; to correct for nightside emission, we estimate the average nightside planet-to-star flux ratio and subtract it from the transit depth. We calculate planet-to-star flux using a NextGen spectrum for the star (interpolated to $T_\mathrm{eff} = 6110$ K), a blackbody for the planet, and a planet-to-star radius ratio of 0.1146. We assume a nightside temperature of 1700 K, which is near the median of the nightside temperatures estimated from the phase variation models (Table\,\ref{table:models}). We also show the uncorrected transit depths. The corrected and uncorrected transit depths are listed in Table\,\ref{table:tspec}. For the uncorrected data, there is an offset between the \HST\ and \Spitzer\ data of more than five atmospheric scale heights.  The uncorrected spectrum is inconsistent with a flat line at $5.3\,\sigma$ confidence, whereas the nightside-corrected spectrum is consistent within $1\,\sigma$.  The corrected spectrum is also consistent with predictions from the $\tau_\mathrm{drag4}$ GCM, which shows water features in the WFC3 bandpass. Future observations with higher precision could reveal these features. 

\begin{deluxetable}{LLLL}
\tablecolumns{4}
\tablewidth{0pt}
\tablecaption{WASP-103b Transmission Spectrum\label{table:tspec}}
\tablehead{
\colhead{Wavelength} & \colhead{$(R_p/R_s)^2$ (\%)} & \colhead{$(R_p/R_s)^2$ (\%)} & \colhead{Error}\\
\colhead{(micron)} & \colhead{($T_\mathrm{n} = 0$ K)} & \colhead{($T_\mathrm{n} = 1700$ K)} & \colhead{(\%)}}
\startdata
1.175 & 1.3178 & 1.3115 & 0.0092 \\
1.225 & 1.3144 & 1.3067 & 0.0087 \\
1.275 & 1.3195 & 1.3103 & 0.0086 \\
1.325 & 1.3454 & 1.3345 & 0.0087 \\
1.375 & 1.3512 & 1.3385 & 0.0089 \\
1.425 & 1.3588 & 1.3441 & 0.0091 \\
1.475 & 1.3364 & 1.3197 & 0.0092 \\
1.525 & 1.3522 & 1.3333 & 0.0093 \\
1.575 & 1.3688 & 1.3476 & 0.0100 \\
1.625 & 1.3549 & 1.3314 & 0.0104 \\
3.6 & 1.4013 & 1.3238 & 0.0328 \\
4.5 & 1.4329 & 1.3554 & 0.0264 \\
\enddata
\tablecomments{Wavelength dependent transit depths, corrected for companion dilution and nightside flux (assuming nightside temperatures of 0 and 1700 K for the second and third columns, respectively). The error corresponds to the 68\% credible interval from an MCMC fit to the transit light curves.}
\end{deluxetable}

\section{Atmospheric Composition and Thermal Structure}
\label{sec:composition}
We characterized the planet's atmospheric composition by fitting 1D models to the phase-resolved emission spectra.  First, we modeled the planet flux as a simple blackbody to estimate the dayside brightness temperature and test for significant absorption or emission features. We then performed a more sophisticated grid-based retrieval to estimate the atmospheric metallicity, carbon-to-oxygen ratio, and thermal structure. We also evaluated the climate based on the best fit \texttt{spiderman} temperature maps.

\subsection{Blackbody Fits}
\label{sec:bbfits}
A blackbody is the simplest model for the planet's thermal emission and provides a useful first evaluation of the atmospheric properties. To calculate the best fit blackbody model, we assumed a planet-to-star radius ratio of 0.1146 and used a NextGen stellar spectrum interpolated to an effective temperature of 6110 K \citep{allard12}.  We calculated the best fit with a least-squares fitting routine. To determine uncertainties on the planet brightness temperature, we performed an MCMC analysis with free parameters for the planet temperature and the stellar $T_\mathrm{eff}$. We used a Gaussian prior on $T_\mathrm{eff}$ of $6110 \pm 160$ K \citep{gillon14}.  

The best-fit blackbodies are shown in Figures \ref{fig:spectra} and \ref{fig:dayside}. The model fits the data fairly well: it is consistent with the data at the $3\,\sigma$ level for 70\% of the orbital phases. However, the \Spitzer\ data have higher brightness temperatures than the WFC3 data at phase 0.5, and lower brightness temperatures at phases $0.8 - 0.9$. These differences suggest the presence of an emission feature on the dayside at \Spitzer\ wavelengths, which transitions to an absorption feature on the nightside, perhaps indicating changes in thermal structure with longitude in the atmosphere. 

We also fit independent blackbody models to the \HST/WFC3 data and each \Spitzer\ channel separately. The resulting brightness temperatures and $1\,\sigma$ uncertainties are listed in Table\,\ref{table:teffs}.  The WFC3 data agree well by a blackbody model at all orbital phases except phase 0.5 (consistent at better than $1.5\,\sigma$). The dayside has higher signal-to-noise than the other orbital phases, thanks to the two secondary eclipse observations from \cite{cartier17}. The more sophisticated grid-based retrieval (described in the next section) provides a better fit to the dayside. 

We note that the uncertainties on the brightness temperatures in
different bandpasses are correlated with each other because they include the
uncertainty on the stellar temperature: i.e., if the stellar temperature
increases, so do the brightness temperatures. To evaluate the significance of
features in the emission spectra, we hold the stellar spectrum fixed in the retrieval analysis.



\begin{deluxetable}{LLLLL}
\tablecolumns{5}
\tablewidth{0pt}
\tablecaption{Phase-resolved Brightness Temperatures \label{table:teffs}}
\tablehead{
\colhead{Orbital Phase} & \colhead{$T_\mathrm{b}$} & \colhead{$T_\mathrm{b}$} &
\colhead{$T_\mathrm{b}$} & \colhead{$\chi^2_\nu$} \\
\colhead{\,} & \colhead{WFC3} & \colhead{Ch. 1} & \colhead{Ch. 2} & \colhead{(9 dof)}}
\startdata
0.06 - 0.15 & 1883 \pm 41 & 1523 \pm 153 & 1589 \pm 105 & 0.7 \\
0.15 - 0.25 & 2208 \pm 33 & 2612 \pm 117 & 2299 \pm 100 & 0.9 \\
0.25 - 0.35 & 2587 \pm 37 & 2926 \pm 114 & 2592 \pm 96 & 1.5 \\
0.35 - 0.44 & 2831 \pm 39 & 2834 \pm 111 & 2976 \pm 93 & 1.5 \\
0.44 - 0.56 & 2933 \pm 41 & 2995 \pm 159 & 3154 \pm 99 & 2.8 \\
0.56 - 0.65 & 2811 \pm 39 & 2631 \pm 110 & 2891 \pm 97 & 2.0 \\
0.65 - 0.75 & 2572 \pm 36 & 2708 \pm 117 & 2421 \pm 100 & 1.3 \\
0.75 - 0.85 & 2263 \pm 33 & 1952 \pm 125 & 1939 \pm 99 & 1.3 \\
0.85 - 0.94 & 1987 \pm 37 & 1594 \pm 145 & 1288 \pm 118 & 0.6 \\
\enddata
\tablecomments{$\chi^2_\nu$ values are for the fits to the WFC3 data only.}
\end{deluxetable}

\subsection{Grid-Based Retrieval}
\label{sec:retrieval}
To infer abundances from the dayside spectrum (phase $0.46 - 0.54$), we use a self-consistent grid-based method (ScCHIMERA) similar to that employed in \cite{arcangeli18, mansfield18}. We generated a grid from one-dimensional forward models of the atmosphere over a broad range of metallicities (M/H), carbon-to-oxygen ratios (C/O), and stellar irradiation ($f$). The $f$ parameter is a scaling factor for the stellar flux at the top of the atmosphere, where $f=1$ corresponds to full heat redistribution and $f=2$ is equivalent to only allowing the dayside to re-radiate.  

At each point in the grid, we compute forward models to determine self-consistent, radiative-convective temperature-pressure (T-P) profiles. We determine the molecular abundances in each atmospheric layer assuming thermochemical equilibrium \citep[calculated with the NASA CEA routine;][]{gordon94}.  We include opacity from the major absorbers expected for a hot Jupiter atmosphere, including H$_2$O, CO, CO$_2$, TiO, VO, FeH, and H$_2$-H$_2$ CIA. Notably, in contrast to most prior atmospheric retrievals for the hottest planets, we also included opacity from H-, which is an important absorber at temperatures above 2500 K \citep{arcangeli18, parmentier18}.  Using these opacities and T-P profiles, we calculated thermal emission spectra over the full grid using the CHIMERA retrieval suite \citep[described in][]{line13a, line14}.  We then explored the grid with an MCMC chain using the \texttt{emcee} package \citep{foremanmackey13} and interpolated in the grid to calculate the likelihood at each model step. The priors were uniform over the ranges $0.2 \le f \le 2.8$, $-1\le \log{\mathrm{M/H}} \le 2.5$, and $-1 \le \log{\mathrm{C/O}} \le 0.95$.  We fit this model to the dayside and nightside emission spectra (phases $\phi = 0.5$ and $0.1$). 

\subsubsection{Dayside Spectrum}
The main characteristics of the dayside emission spectrum are: (1) it is blackbody-like at WFC3 wavelengths, and (2) in the \Spitzer\ bands, the planet-to-star flux is significantly higher than predicted for the best fit blackbody, indicating an emission feature.  The best-fit spectrum reproduces these data fairly well, with $\chi^2_\nu$ = 1.77 (for 9 degrees of freedom). The largest contribution to the $\chi^2$ value is the $4.5\,\mu$m eclipse depth, which is larger than the best fit model prediction by $2.9\,\sigma$. When the $4.5\,\mu$m point is removed, the fit has $\chi^2_\nu = 1.17$ (8 degrees of freedom).  The best fit model has a moderately enhanced metallicity ($23\times$ solar), carbon-to-oxygen equal to 0.76, poor heat redistribution, and a thermal inversion (temperature increasing with altitude).

\begin{figure}
\includegraphics[width = 0.5\textwidth]{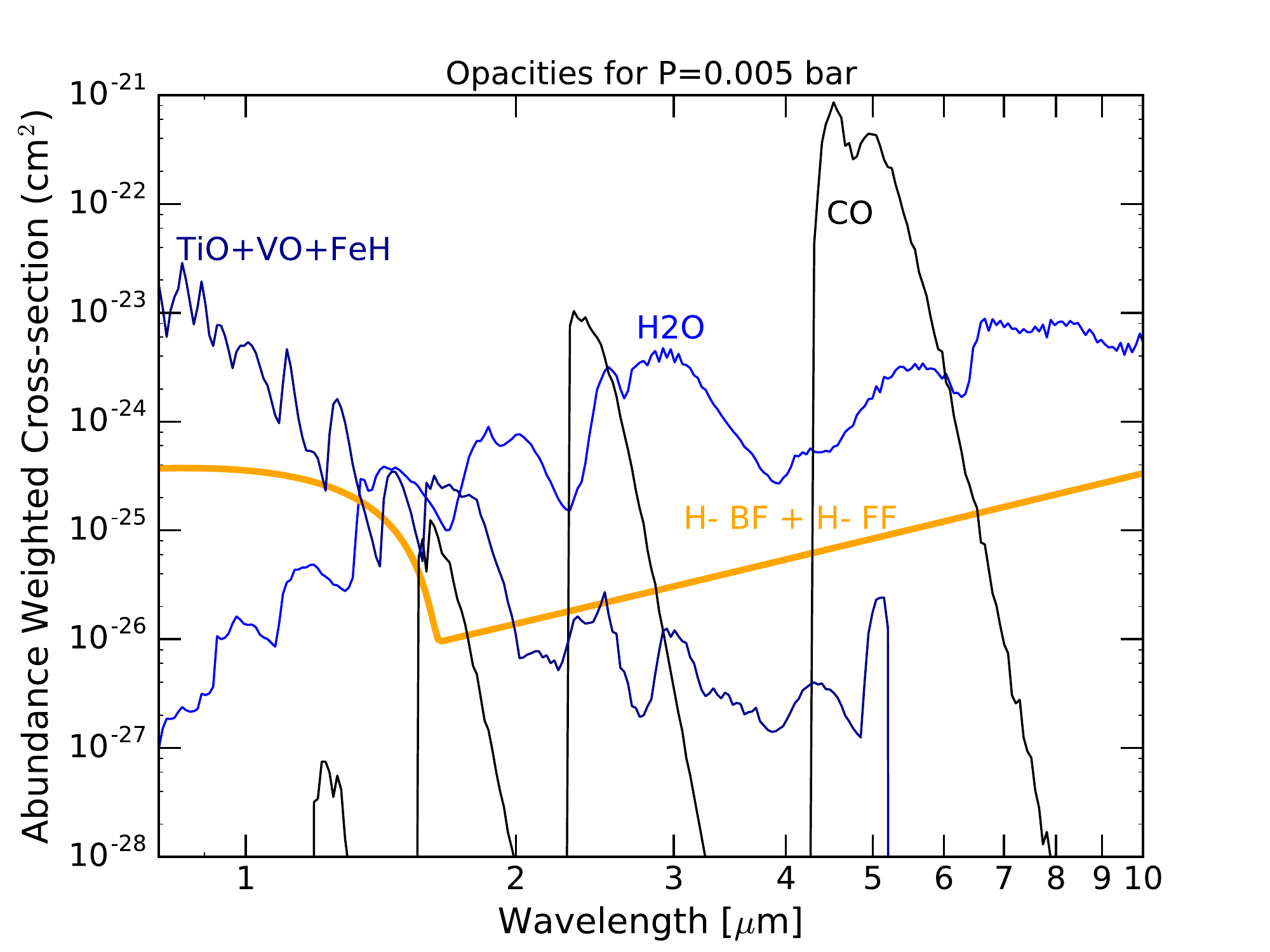}
\caption{Abundance weighted absorption cross-sections illustrating the important opacity sources at the photospheric pressure and temperature (5 mbar, 3036 K). The strong CO feature at 4.5\,$\mu$m contributes to the high planet-to-star flux at that wavelength.  Water, hydrides/oxides, and the H- bound-free opacities all play a role in shaping the \HST/WFC3 spectrum.}
\label{fig:opacities}
\end{figure}

\begin{figure}
\includegraphics[width = 0.5\textwidth]{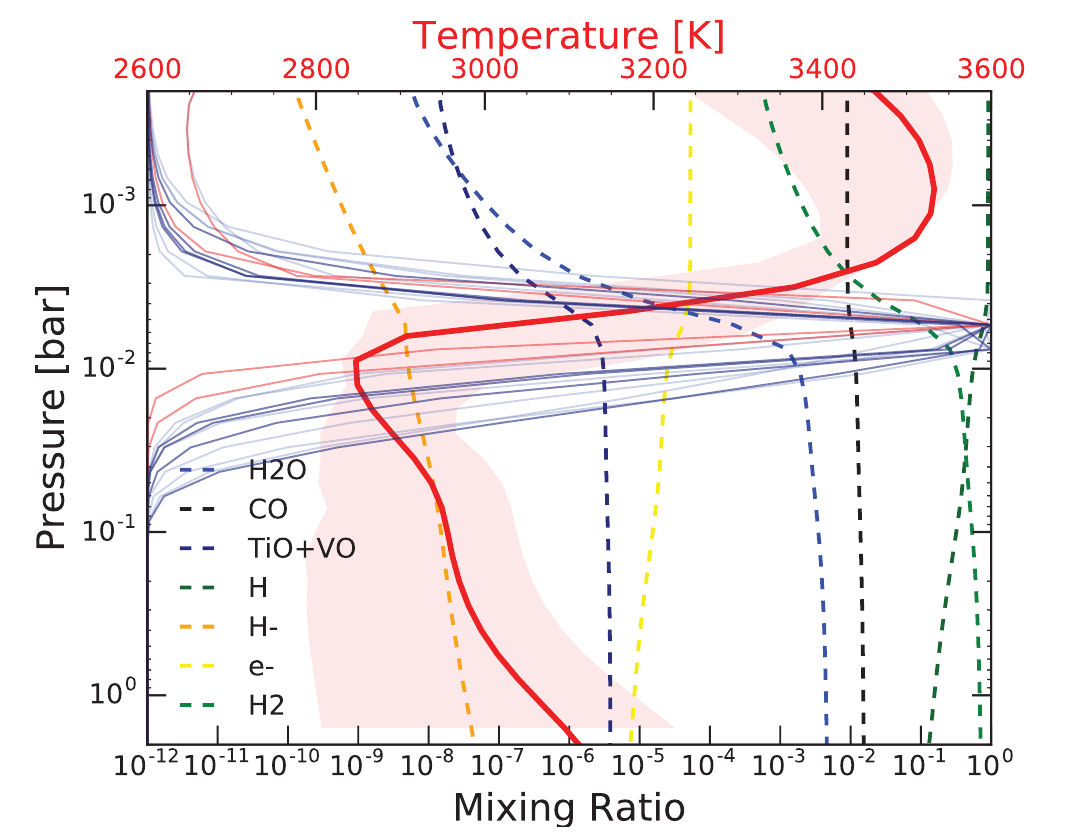}
\caption{Summary of the 1D self-consistent model atmosphere fits to the dayside emission spectrum.  The temperature-pressure profile (top axis) is indicated by the 1-sigma spread of 500 randomly drawn T-P profiles from the posterior (light red) and a representative fit, with $f$=0.4, [M/H]=1.5, and C/O=0.7 (dark red).  The normalized thermal emission contribution functions for the \Spitzer\ points are shown in solid red, the WFC3 in-water band ($1.33-1.48\,\mu$m) in dark blue, and WFC3 out-of-water-band in light blue.  The observations probe between $\sim0.01$ and 0.001 bar, just above the tropopause region of the atmosphere where the temperature is increasing.  The dashed curves are thermochemical equilibrium mixing ratios for important absorbers computed along the representative fit's self-consistent T-P profile.  Note the rapid dissociation of water above the $\sim10$ mbar level where the inversion begins.}
\label{fig:summary}
\end{figure}

Figure\,\ref{fig:opacities} shows the opacity contributions of key absorbers for the best fit model.  In the optical (which we do not observe directly), there is strong absorption by TiO, VO, and FeH.  In the near-infrared, H$_2$O, H-, and hydrides/oxides all contribute to the opacity, leading to nearly constant opacity over the WFC3 wavelength range.  In cooler atmospheres, water is the dominant absorber over this bandpass \citep[e.g.][]{kreidberg14b, line16}, but in WASP-103b, H$_2$O is partially dissociated in the photosphere, leading to a drop in abundance by a factor of $\sim10$ (see Figure\,\ref{fig:summary}). Water also has intrinsically weaker features at high temperature \citep[e.g.][]{tinetti12}.  On top of this, there is significant H- opacity from single H atoms bound with free electrons, which fills in the opacity at wavelengths shorter than $1.5\,\mu$m. Finally, the sharp vertical gradient in water abundance results in water becoming optically thick over a very narrow range in pressure, where temperature is nearly constant.  Taken together, all these factors add up to produce a nearly featureless spectrum from $1.1 - 1.7\,\mu$m. Finally, in the infrared the dominant absorber is CO, which produces the emission feature at \Spitzer\ wavelengths.  

Figure\,\ref{fig:summary} shows a summary of the temperature-pressure profile and abundances for the best fit model. The T-P profile is inverted, with temperature increasing from 2800 to 3500 K over the pressure range $10^{-2} - 10^{-3}$ bar.  The thermal inversion is probably driven by absorption of optical light by oxides and hydrides in the upper atmosphere and the absence of cooling by water molecules (which have dissociated).  The observations are sensitive to pressures of $\sim0.01 - 0.001$ bar, which spans the tropopause, where temperature begins to increase and the water abundance drops by more than an order of magnitude.

\begin{figure}
\includegraphics[width = 0.5\textwidth]{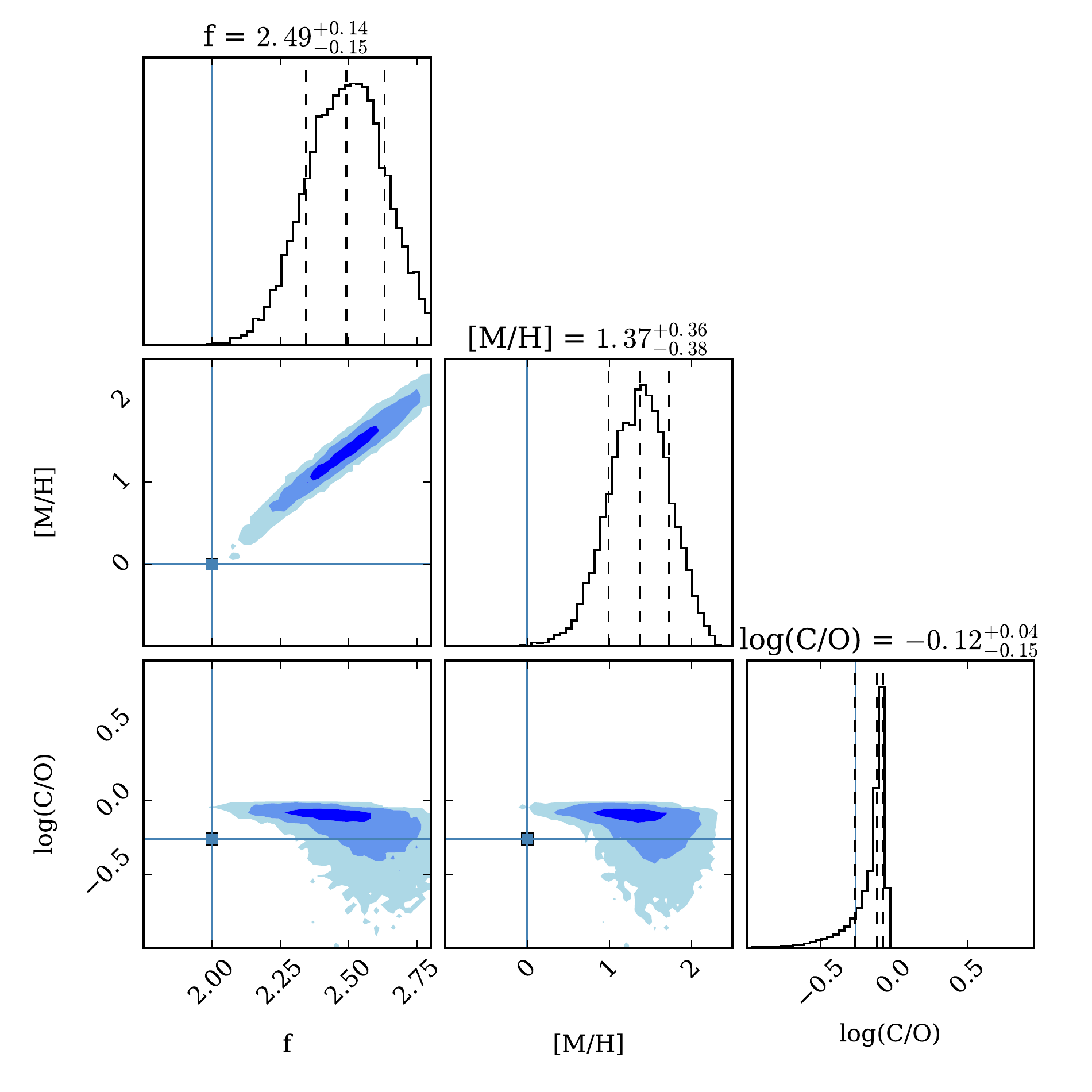}
\caption{Posterior distributions for WASP-103b's atmospheric heat redistribution, metallicity, and C/O, from a grid-based fit to the dayside emission spectrum. The histograms on the diagonal show the marginalized distribution of each parameter, with dashed lines indicating the median and surrounding 68\% credible interval. The blue lines correspond to solar metallicity (1) and C/O (0.54). The 2D histograms mark the 1, 2, and 3$\,\sigma$ credible regions in dark, medium, and light blue, respectively.}
\label{fig:composition}
\end{figure}

In Figure\,\ref{fig:composition}, we show the posterior distributions from the grid retrieval.  We infer a range in metallicity of $23^{+29}_{-13}\times$ solar, somewhat higher than expected based on Jupiter's metal enrichment \citep[$3-5\times$ solar;][]{wong04} and the trend toward decreasing metallicity with increasing planet mass observed for the Solar System and exoplanets \cite[e.g.][]{kreidberg14b}.  The metallicity is super-solar at $>3\,\sigma$ confidence. However, planet population synthesis models predict some scatter in atmospheric metallicity. Planets near WASP-103b's mass ($1.5\,M_\mathrm{Jup}$) are expected to have metallicities ranging from roughly $1-10\times$ solar \citep{fortney13, mordasini16}. Our result for WASP-103b lies on the upper end of this range, and may be indicative of intrinsic scatter in the mass-atmospheric metallicity relation. 

The retrieved C/O is consistent with solar, with a $1\sigma$ confidence interval of $0.54 - 0.85$. We infer an upper limit on C/O of 0.9 at $3\,\sigma$ confidence, driven by the fact that the atmospheric chemistry is expected to change dramatically when C/O exceeds unity. For a carbon-rich composition, the equilibrium abundance of methane relative to CO increases by orders of magnitude compared to an oxygen-rich  composition \citep[e.g.][]{madhusudhan11}. Our \Spitzer\ eclipse depths are sensitive to the relative abundance of these species, so we can confidently rule out a carbon-rich composition despite the lack of spectrally resolved features (assuming the atmosphere is in chemical equilibrium).

We infer a heat redistribution $f = 2.49^{+0.14}_{-0.15}$. An $f$ parameter of unity represents isotropic heat distribution, whereas $f = 2$ corresponds to dayside emission only. We estimate $f > 2$, indicating a thermal inversion and likely inefficient transport of heat to the nightside.  The heat redistribution is strongly correlated with atmospheric metallicity because increasing metallicity  shifts the T-P profile to lower pressures, resulting in hotter temperatures at a given pressure level (equivalent to less efficient heat redistribution). 

There are several caveats in our analysis which are important to note:
\begin{my_itemize}
    \item {The best fit model is not a perfect fit to the data (with $\chi_\nu = 1.77$ for 9 degrees of freedom), so the uncertainties produced by the MCMC may be underestimated.} 
    \item {The inferred C/O and metallicity are highly sensitive to the planet-to-star flux at \Spitzer\ $4.5\,\mu$m, which is the worst fit data point. To fit this data point, the model favors super-solar metallicities and C/O, which drive up the CO abundance (the dominant absorber at $4.5\,\mu$m).}
    \item{The \Spitzer\ $4.5\,\mu$m data is from broadband photometry, so the inferred CO feature is not spectrally resolved. It is possible that unknown absorbers or disequilibrium chemistry affect the $4.5\,\mu$m planet-to-star flux, but we cannot uniquely identify these features in our spectrum.}
\end{my_itemize}
We therefore caution against over-interpreting these results until wider spectral coverage is available.


\subsubsection{Nightside Spectrum}
We also fit the nightside spectrum (phase $0.1$) with the grid-based retrieval.
The best fit spectrum has a non-inverted temperature pressure profile.  At
1\,$\sigma$ confidence, the metallicity is $15 - 240\times$ solar and the C/O is
unbounded over the full prior range. The atmospheric composition is consistent
with results from the dayside spectrum. 

This agreement is an encouraging sanity check; however, there are several model assumptions that may result in artificially tight constraints on the atmospheric properties on the nightside.  One challenge in modeling the nightside spectrum is that the physical processes shaping the T-P profile are unknown.  Our model assumes a scaled stellar irradiation at the top of the atmosphere, but in reality, the heat source is advection from the dayside. Another caveat is that the model is not self-consistent: the energy leaving the dayside is not constrained to equal the energy entering the nightside.  Further work is needed to develop a fully self-consistent 2-D retrieval method for phase curve observations. 

As a test, we also calculated the difference in brightness temperature
between the \HST\ and \Spitzer\ 4.5 $\mu$m data (reported in
Table\,\ref{table:teffs}) for both nightside phases ($\phi = 0.1$ and 0.9). At
phase 0.1 and 0.9, the \Spitzer\ temperature is lower by 2.7 and 5.6\,$\sigma$,
respectively.  These values are a lower limit to the significance, because the
brightness temperatures noted in the table also include the uncertainty in the
stellar $T_\mathrm{eff}$ (which increases the uncertainty on the absolute planet
temperature but not the relative temperatures that are relevant for this
calculation). The drop in brightness temperature is more significant at phase 0.9 than it is at 0.1, providing further evidence in addition to the phase 0.1 retrieval that the nightside temperature pressure profile is not inverted.

\subsection{Climate}
We fit three different models to characterize the planet's climate: a two-temperature map, the physically-motivated kinematic model of \cite{zhang17}, and a spherical harmonic map. We also fit the thermal phase variation with a sinusoid, which can be inverted to map the climate \citep{cowan08, cowan17}.  All of the models provide reasonable fits to the data, with $\chi^2_\nu$ near unity, but they yield significantly different temperature maps. Table\,\ref{table:models} lists the best fit minimum and maximum temperatures, as well as the mean day- and nightside temperatures. We also list the  information criterion (BIC) values for the fits \citep[a $\Delta$BIC value greater than 10 constitutes strong evidence against a given model;][]{kass95}.

The spherical harmonics map generally fits the data the best. It has a lower BIC value than all the other models for the broadband WFC3 and \Spitzer\,$3.6\,\mu$m phase curves. For the $4.5\,\mu$m phase curve, the two temperature map provides the best fit, but it only lowers the BIC value by 2 relative to the spherical harmonics map, which is not a statistically significant improvement \citep{kass95}.  The spherical harmonics model yields a mean dayside temperature near 2700 K, whereas the nightside is closer to 1800 K, in good agreement with the blackbody fits to the phase-resolved spectra (see \S\,\ref{sec:bbfits}).  The other models produce more extreme day-night temperature gradients.  Between the models, there are substantial differences in the minimum and maximum temperatures (sometimes over 1000 K), whereas the day and nightside means are in better agreement (generally matching to within 250 K). This behavior is not surprising: a wide range of temperature gradients can yield similar average temperatures when integrated over the disk of the planet.  

\section{Comparison with GCMs}
\label{sec:gcm}
To explore the three-dimensional effects of atmospheric dynamics, we ran several GCMs to compare with the measured phase-resolved spectra.  We simulated the atmospheric circulation and thermal structure of the planet using the combined SPARC/MITgcm model~\citep{showman09}. The model solves the primitive equations in spherical geometry using the MITgcm~\citep{Adcroft2004} and the radiative transfer equations using a state-of-the-art one dimensional radiative transfer model~\citep{Marley1999}. The code represents the opacities as correlated-k tables based on the line-by-line opacities described in \citet{Visscher2006,freedman14}. Our fiducial model assumes a solar composition with elemental abundances of \citet{Lodders2002a} and the chemical equilibrium gas phase composition from \citet{Visscher2006}. These calculations take into account the presence of $\rm H^-$ opacities and the effect of molecular dissociation on the abundances. We used a timestep of 25s, ran the simulations for 300 days, and averaged all quantities over the last 100 days. Overall, our setup is the same as described in~\citet{parmentier16} but uses the specific parameters for the WASP-103 system.

Several physical processes can reduce the ability of the atmosphere to transport heat efficiently through advection and change the overall circulation pattern. Among them, ohmic drag is though to be an important phenomenon in the ionized environment of extremely hot hot Jupiters~\citep{perna10}. We parametrize this effect as a Rayleigh drag with a drag constant $\tau_{drag}$ constant with pressure~\citep{showman13}. Varying $\tau_{drag}$  from large values (i.e. weak drag) to small values (i.e. strong drag), the atmospheric circulation is expected to shift from a jet-dominated regime to a more axisymmetric circulation pattern going from the substellar to anti stellar point. Moderate drag timescales are expected to change the circulation pattern and thus reduce significantly the shift of the hottest point of the atmosphere whereas short drag timescales are also expected to change strength of the winds and thus the atmospheric day/night contrast~\citep{komacek16,komacek17}.  Although Rayleigh drag is an incomplete representation of the complex magneto-hydrodynamic effects expected in these atmospheres~\citep{batygin13,rogers14,rogers14a,rogers17}, it nonetheless provides an estimate of the strength of the drag mechanism necessary to match the observations~\citep{komacek17,koll18,parmentier17}. 

\begin{figure}
\includegraphics[width = 0.5\textwidth]{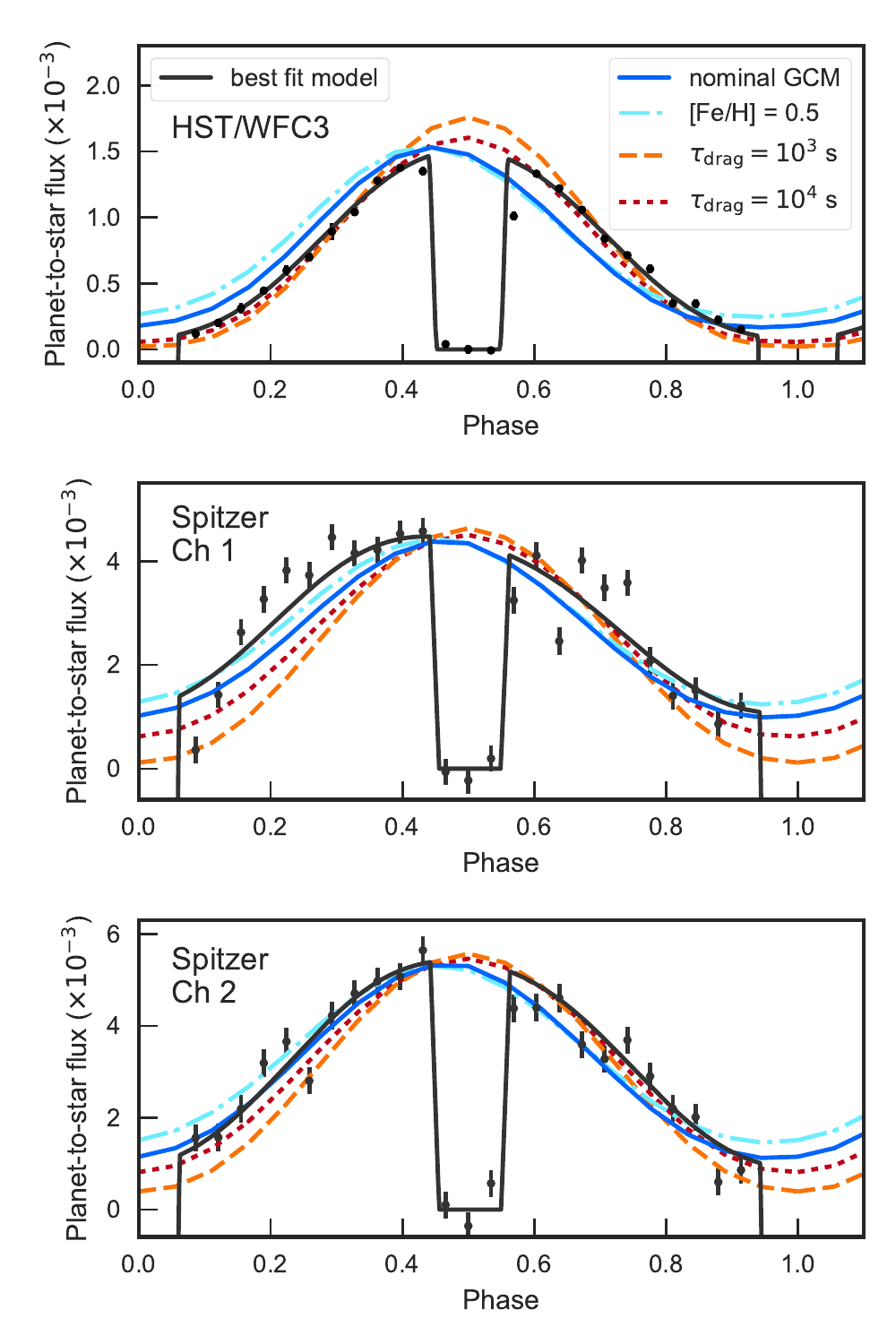}
\caption{GCM predictions (colored lines) compared to the best fit spherical harmonics model for the WFC3 white light, \Spitzer\ 3.6, and \Spitzer\ 4.5 micron phase curves (black lines, top to bottom). The nominal model is solar composition, cloud-free, with TiO/VO opacity and no drag. The models are corrected for the predicted ellipsoidal variability of the planet.}
\label{fig:gcmcomparison}
\end{figure}

Our nominal GCM was a cloud-free, solar composition atmosphere with TiO/VO
opacity and no added drag.  Each GCM run is computationally intensive, so we ran
a small number of additional models to see which parameters had the largest
effect on the planet spectrum. We changed model parameters one at a time,
considering cases with enhanced metallicity ([M/H] = 0.5), no TiO/VO, and added
atmospheric drag with timescales $\tau_\mathrm{drag} = 10^3$ and $10^4$ s, which
we label $\tau_\mathrm{drag3}$ and $\tau_\mathrm{drag4}$, respectively.  The GCM
results are shown in Figure\,\ref{fig:gcmcomparison}. To assess how well the GCM predictions reproduce the data, we calculated the amplitude and hot spot offset for all the models (listed in Table\,\ref{table:amps_offsets}). The small observed hotspot offsets ($-0.3 - 2.0$ degrees) are best reproduced by the $\tau_\mathrm{drag4}$ model, which has a smaller offset than the drag-free GCMs due to changes in wind pattern. In the drag models, the winds shift from a substellar to an antistellar flow rather than an equatorial jet.  The $\tau_\mathrm{drag4}$ model also provides the match the observed phase curve amplitudes. 

\begin{figure}
\includegraphics[width = 0.5\textwidth]{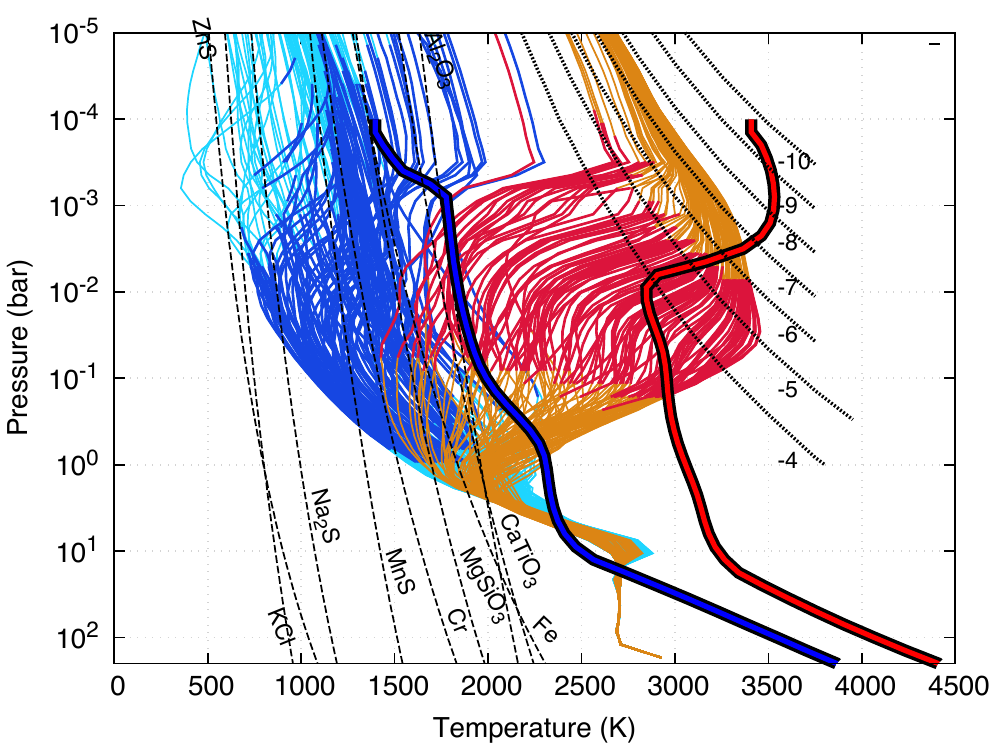}
\caption{TP profiles from the GCM and 1D retrieval, compared to condensation curves of potential cloud species. The thin blue/cyan lines correspond to a sample of randomly drawn nightside TP profiles from the GCM, and the thin red/orange lines correspond to dayside TP profiles. The darker colors (blue/red) indicate the extent of the contribution function (encompassing 80\% of the emitted flux). The thick red and blue lines are the best fit 1D TP profiles for the phase 0.5 (dayside average) and 0.1 (nightside average). The dashed lines are condensation curves for a range of possible cloud species.  The dotted lines correspond to regions of constant H$_2$O abundance, with numbers indicating the log10(H$_2$O volume mixing ratio).}
\label{fig:TP}
\end{figure}

We also compared the TP profiles from the $\tau_\mathrm{drag4}$ GCM to cloud condensation curves and the best fit radiative-convective equilibrium models from the 1D retrieval (Figure\,\ref{fig:TP}).  For the dayside photosphere, the TP profile slope and absolute temperature are in rough agreement between the 1D best fit and the GCM. At higher pressures, the GCM is systematically cooler, which is likely due to the effect of atmospheric circulation (at these pressures the GCM mixes the temperature planet-wide). At lower pressures, the GCM is also cooler than the 1D fits, which may be due to metallicity differences between the models. The GCM has solar metallicity, whereas the best fit 1D model has $[M/H] \sim 1$. Higher metallicity compositions have larger TiO/H$_2$O ratios, and since the pressure dependence of TiO dissociation is not as strong as for water dissociation, we expect stronger inversions for higher metallicity atmospheres \citep{parmentier18}.  On the nightside, we also find that the GCM is cooler than the 1D models.  While in the 1D model the day-to-night redistribution is fitted to the data, it is not a tunable parameter in the 3D GCM. There are several physical processes not included in the GCM that could contribute to a hotter nightside, including shocks, longitude-dependent drag, and latent heat released from H2 recombination \citep{bell18}.  The best fit nightside TP profile is hotter than the condensation curves through most of the photosphere, suggesting that the observable atmosphere is relatively free of clouds. This prediction could be tested with longer wavelength phase curve observations.

We also compared the GCM output to temperature maps retrieved with \texttt{spiderman}. Figure\,\ref{fig:model_comparison} shows the 0.1 bar temperature map for the $\tau_\mathrm{drag4}$ GCM compared to the best fit models. At this pressure, the GCM has minimum and maximum temperatures of $920$ and $3360$ K. The temperature gradient from the dayside to the terminator is intermediate between the kinematic and spherical harmonics models. The GCM predicts a cooler nightside than all models except the two-temperature model.  We note that none of the models are perfect: there is degeneracy in the \texttt{spiderman} maps, with large differences in climate producing reasonably good fits to the phase curves (see Table\,\ref{table:models}), whereas the GCM is too cold on the nightside.  Robustly constraining the climate will require more sophisticated GCMs and higher precision phase curves/eclipse mapping \citep[e.g.][]{dewit12}.

\begin{figure*}
	\includegraphics[width = 1.0\textwidth]{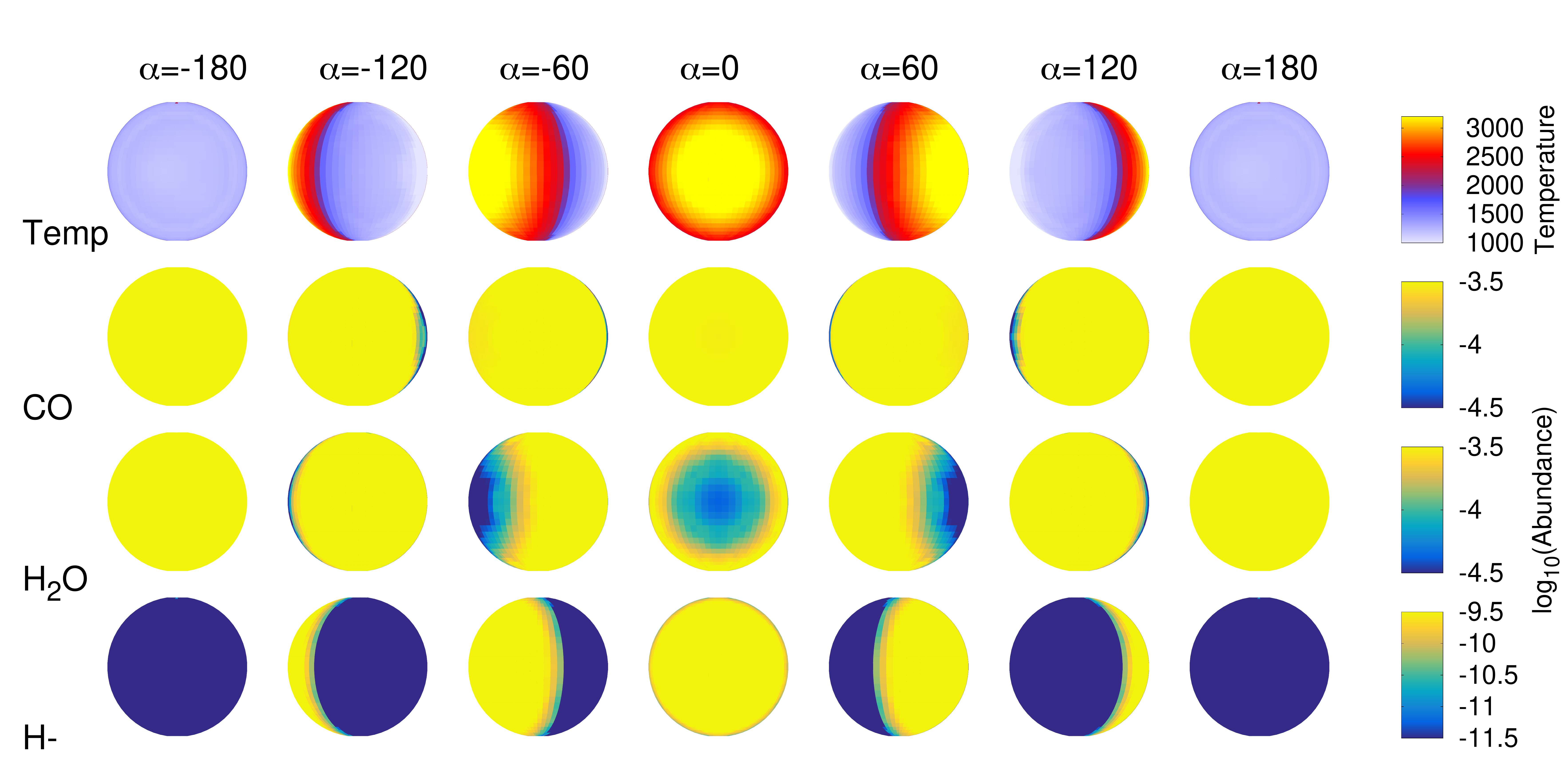}
	\caption{Row (1): photospheric temperatures for the $\tau_\mathrm{drag4}$ GCM for different viewing geometries.  The orbital phase $\alpha=0$ corresponds to secondary eclipse (when the substellar point faces Earth) and $\alpha=180$ corresponds to transit (when the antistellar point faces Earth). Rows (2-4) show the abundances of CO, H$_2$O, and H-. At dayside temperatures, water dissociates, so the photospheric water abundance drops by $\sim10$ and the H- abundance increases by $\sim100$. By contrast, the CO abundance is uniform throughout the photosphere.}
\label{fig:GCMabundance}
\end{figure*}

The GCMs also provide insight into what molecules are present in which parts of the atmosphere. As discussed in \S\,\ref{sec:composition}, water dissociation and H- opacity are needed to explain the dayside emission spectrum.  Figure\,\ref{fig:GCMabundance} shows the photospheric abundances of H$_2$O and H- compared to the predicted temperature for the $\tau_\mathrm{drag4}$ GCM. The water abundance drops by $\sim10$ at the substellar point, and the H- opacity increases by $\sim100$.  By contrast, CO remains intact throughout the atmosphere.  Our observations are not precise enough to detect water features on the nightside of the planet (see \S\,\ref{sec:comparison}), but future high precision data may be sensitive to these features, and will help constrain the strength of horizontal transport in the atmosphere \citep{agundez14}.  

\subsection{Constraints on the Planet's Magnetic Field}
We show in this section that the small observed hotspot offset in the phase curves is best fit by a GCM that includes Rayleigh drag with a timescale $\tau_\mathrm{drag} = 10^4$\,s. This observation gives rise to the question: what magnetic field strength on the planet can produce drag with this timescale?  Previous efforts to characterize exoplanet magnetic fields have mainly focused on magnetic interaction between the planet and its host star \citep[e.g.][and references therein]{wright15} and planetary radio emission \citep{griessmeier15}.  A complementary approach is to study the effect of the magnetic field on the planet's atmospheric dynamics.

Here we make a simple order-of-magnitude estimate for the magnetic field required to produce a drag timescale of order $10^4$\,s. We first computed the free electron abundance due to ionized metals for the $\tau_\mathrm{drag4}$ GCM. Using the analytic expression from \cite{perna10} (Equation 12), we estimated a drag timescale at the substellar point of $\tau_\mathrm{drag}  = 2\times10^4/B^2$ s, where $B$ is the magnetic field in Gauss. We assumed a temperature and pressure of 3359 K and 0.11 bar, 
and that the magnetic field is perpendicular to the flow. To reach a drag timescale of $10^4$\,s, a magnetic field stronger than $\sim1$ Gauss is required, comparable to Jupiter's magnetic field strength of 5-10 Gauss \citep{bagenal04}.  To confirm this intriguing result, more detailed study is warranted, including a full magnetic hydrodynamic simulation of the atmospheric dynamics \citep[e.g.][]{rogers17} and accounting for the possibility of complex magnetic field structure due to interactions between the magnetic fields of the planet and the star.

\section{Comparison with Brown Dwarfs and Directly Imaged Companions}
\label{sec:comparison}
WASP-103b is so highly irradiated that its photospheric temperature ($2000 - 3000$ K) is comparable to that of low mass stars. However, the planet's other properties (surface gravity, rotation rate, irradiation) are different. To explore the effects of varying these parameters, we selected spectra from WASP-103b at three orbital phases: dayside ($\phi = 0.5$), quadrature ($\phi = 0.25$), and nightside ($\phi = 0.1$), and compared them to brown dwarfs and young directly imaged companions with comparable brightness temperatures.

We also used three brown dwarfs/low mass stars for comparison. We chose the field sources: 2MASS J13204427+0409045, (1320+0409) an optical L3, 2MASS J04285096-2253227 (0428-2253) an optical L0.5 and 2MASS J00034227-2822410 (0003-2822) an optical M8 \citep[see][]{filippazzo15}.  We then used all currently available photometric, astrometric, and spectroscopic data for each source to evaluate fundamental parameters such as mass, $T_\mathrm{eff}$ and log $g$ \citep{filippazzo15, faherty16} and create flux-calibrated spectral energy distributions.  For 1320+0409 we used SDSS, WISE and 2MASS photometry along with the optical spectrum from \cite{reid08}, the near infrared spectrum from \cite{bardalez14}, and the parallax reported in \cite{2012ApJ...752...56F}. For 0428-2253 we used 2MASS, DENIS and WISE photometry along with the optical spectrum from \cite{2003A&A...403..929K} the near infrared spectrum from \cite{bardalez14} and the parallax reported in \cite{2014AJ....147...94D}, and for 0003-2822 we used 2MASS and WISE photometry along with the optical spectrum from \cite{2007AJ....133..439C} the near infrared spectrum from \cite{2018AJ....155...34C}, and the parallax reported in \cite{2010AJ....139..176F}.  Coincidentally, both 1320+0409 and 0003-2822 are widely separated ($> 2000$ AU) companions to K2 and G8 stars respectively.  All data was gathered from the Brown Dwarfs in New York City (BDNYC) database \citep{filippazzo15}\footnote{The BDNYC Database: \url{http://database.bdnyc.org/}}.  At the assumed field ages of each source, 0003-2822 would be above the nuclear burning boundary (star) while 1320+0409 would be below (brown dwarf).  0428-2253 is likely a star but at a slightly younger field age could be a brown dwarf.

The directly imaged spectra are for the sources CD-35 2722, USco 1610-1913B, and TWA 22A and are taken from \cite{wahhaj11, aller13, bonnefoy14}. They are young objects (aged 10 - 100 Myr), with lower surface gravities than brown dwarfs of comparable temperature. They also have gravitationally bound companions over a wide range of separations ($67\pm4$, $840\pm90$, and $1.8\pm0.1$ AU, respectively). The sources are calibrated in absolute flux using published H-band photometry \citep{wahhaj11, aller13, bonnefoy09} and distances \citep{gaia16, teixeira09}, a flux-calibrated spectrum of Vega \citep{1985A&A...151..399M, 1985IAUS..111..225H}, and the corresponding filter passbands.  

\begin{figure*}
\includegraphics[width = 1.0\textwidth]{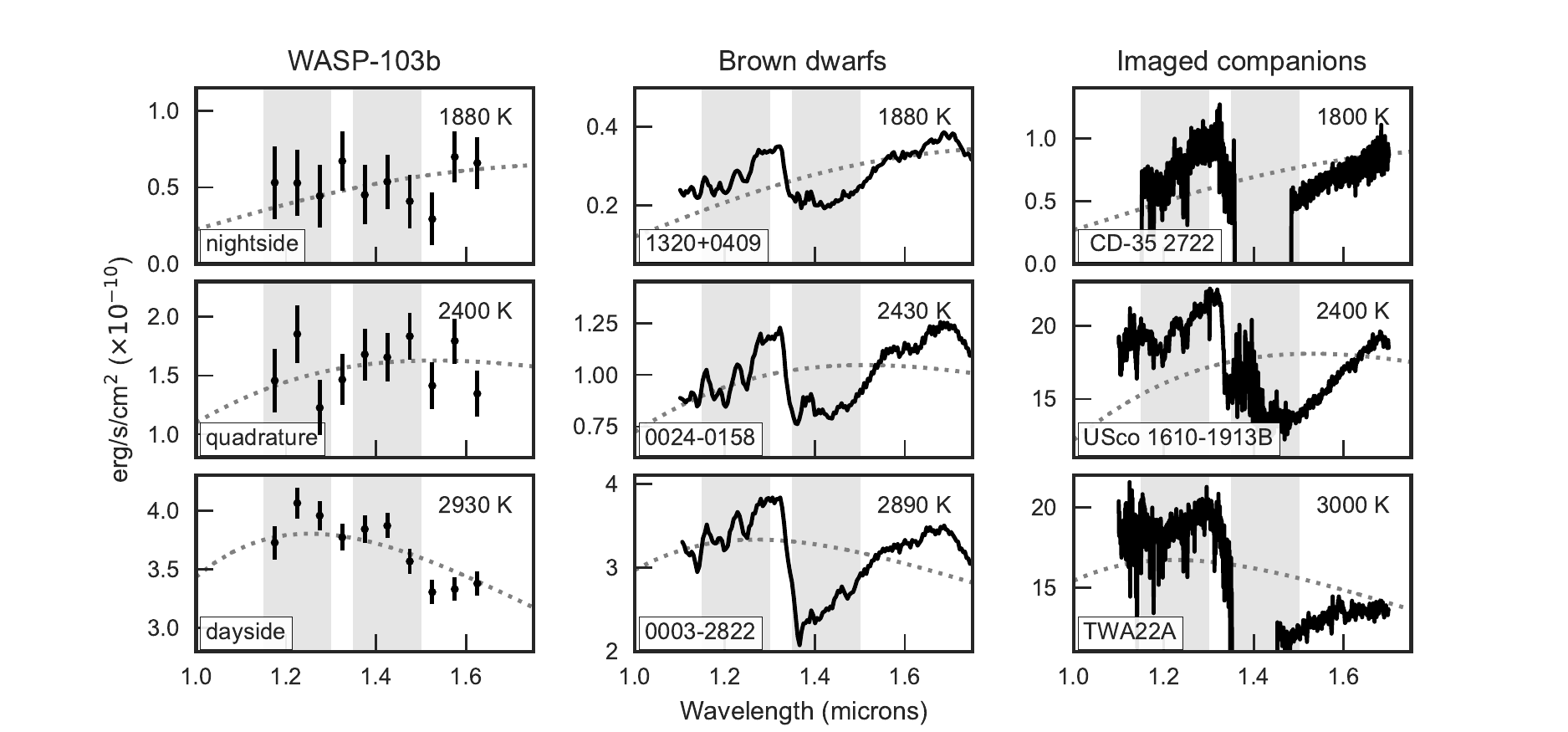}
\caption{Flux-calibrated spectra for WASP-103b (left column) compared to
observed spectra for brown dwarfs (middle) and directly imaged companions
(right), assuming a distance of 10 pc. The WASP-103b spectra are from the
nightside (phase 0.1, top row), quadrature (phase 0.25, middle) and the dayside
(phase 0.5, bottom). Each row shows spectra from objects of comparable
temperature. The dotted gray line corresponds to the best fit blackbody.
Effective temperatures are listed in the upper right corners. The wavelength
bins used to calculate the water feature amplitude $A_1$ are shaded in gray.
Note that the panels do not have the same y-scale, since the objects have
different radii which leads to large variation in absolute flux.}
\label{fig:planetstarcomparison}
\end{figure*}

The system properties for all the objects are summarized in Table\,\ref{table:sources}.  Figure\,\ref{fig:planetstarcomparison} shows the flux-calibrated spectra (assuming a distance of 10 pc).  We compared the spectra over the wavelength range $1.1 - 1.7\mu$m. The most prominent spectral features over this range are expected to come from water, which has a forest of absorption lines near 1.4\,$\mu$m. Spectra for the brown dwarfs and imaged companions have noticeable features in the water band, whereas WASP-103b does not. To quantitatively compare the water feature amplitude for different objects, we define an amplitude $A = (F_{1,2} - F_{3,4})/F_{1,2}$, where $F_{1,2}$ is the weighted mean flux in a wavelength bin $\lambda_1 - \lambda_2$. We calculated the water feature amplitude for two choices of wavelength bins. For the first, $A_1$, we considered data in and out of the water band, with $\lambda_{1,2,3,4} = 1.15, 1.3, 1.35, 1.5\,\mu$m. The ground-based direct imaging data do not span this entire wavelength range, so we also define an amplitude $A_2$ with $\lambda_{1,2,3,4} = 1.2, 1.35, 1.5, 1.65\,\mu$m. The estimated amplitudes and uncertainties are listed in Table\,\ref{table:sources}. We note that a number of indices have been defined to characterize features in brown dwarf spectra, and these indices have revealed trends in the amplitude of a range of spectral features (water, sodium, potassium, VO, and FeH) with surface gravity and effective temperature \citep{reid01,geballe02,mclean03}. However, the WASP-103b data do not have high enough signal-to-noise or spectral resolution to meaningfully compare these indices. Instead we use a broader bandpass to address the simpler question: are the spectra with the same temperaratures consistent with each other?
 
We find that the $A_1$ and $A_2$ values are significantly lower for WASP-103b at dayside and quadrature than for brown dwarfs and imaged companions of similar temperature. WASP-103b typically has $A_1$ and $A_2$ consistent with zero, indicating no water absorption (in agreement with the analysis in \S\,\ref{sec:results} that showed water is depleted in the photosphere). By contrast, the brown dwarfs and young companions have significant water features, with drops in flux of about 20\% in the water band. This is not surprising: stars in the temperature range $2000 - 3000\,\mathrm{K}$ are well known to have prominent water features \citep{kirkpatrick93}.  Based on the grid retrieval of WASP-103b's atmospheric composition, there are several reasons WASP-103b would exhibit different behavior at the same temperature. WASP-103b is irradiated from above rather than below, changing the shape of the temperature-pressure profile. In addition, WASP-103b also has much lower surface gravity (log$g$ = 3.2 versus 4-5 for stars), which pushes the photosphere to lower pressures, where water dissociates more readily \citep{arcangeli18}. These factors are not relevant on the nightside, and 3D models predict that WASP-103b has nightside water absorption features; however, the current data are not precise enough to distinguish between a blackbody spectrum versus water features like those seen in the other objects. 

\begin{deluxetable*}{llCCCC}
	\tablecolumns{6}
	\tablewidth{0pt}
	\tablecaption{Source Properties \label{table:sources}}
	\tablehead{
	\colhead{\,} & \colhead{Object} & \colhead{$T_\mathrm{eff}$ (K)} & \colhead{log\,$g$ (cgs)} & \colhead{H$_2$O A$_1$} & \colhead{H$_2$O A$_2$}}
		\startdata
		Hot Jupiter & W103b night & 1880 \pm 40 & 3.2 \pm 0.04 & 0.07 \pm 1.8\mathrm{e}-01 & -0.00 \pm 1.6\mathrm{e}-01 \\
		\, & W103b quadrature & 2400 \pm 40 & 3.2 \pm 0.04 & -0.14 \pm 7.8\mathrm{e}-02 & -0.01 \pm 6.7\mathrm{e}-02 \\
		\, & W103b dayside & 2930 \pm 40 & 3.2 \pm 0.04 & 0.04 \pm 1.4\mathrm{e}-02 & 0.15 \pm 1.2\mathrm{e}-02 \\
		Brown Dwarf & 2MASS J1320+0409 & 1880 \pm 70 & 5.19 \pm 0.16 & 0.21 \pm 6.3\mathrm{e}-04 & -0.06 \pm 5.0\mathrm{e}-04 \\
		\, & 2MASS J0428-2253 & 2430 \pm 80 & 5.22 \pm 0.09 & 0.16 \pm 1.2\mathrm{e}-04 & -0.03 \pm 1.0\mathrm{e}-04 \\
		\, & 2MASS J0003-2822 & 2890 \pm 80 & 5.18 \pm 0.04 & 0.26 \pm 1.3\mathrm{e}-04 & 0.10 \pm 1.1\mathrm{e}-04 \\
		Imaged Companion & CD-35 2722 & 1800 \pm 100 & 4.5 \pm 0.5 & - & 0.15 \pm 1.0\mathrm{e}-05 \\
		\, & USco 1610-1913B & 2400 \pm 150 & - & 0.27 \pm 2.0\mathrm{e}-04 & 0.19 \pm 2.2\mathrm{e}-04 \\
		\, & TWA 22A & 3000 \pm 100 & 4.5 \pm 0.5 & - & 0.29 \pm 2.0\mathrm{e}-05 \\
		\enddata
	\end{deluxetable*}

\section{Summary}
\label{sec:summary}
We observed thermal phase curves of the hot Jupiter WASP-103b measured with \HST/WFC3 time series spectroscopy ($1.15 - 1.65\,\mu$m) and \Spitzer/IRAC broadband photometry ($3.6$ and $4.6\,\mu$m bands). Here we summarize our conclusions about the atmosphere based on these measurements.

\begin{itemize}
	\item{The dayside planet-to-star flux is $0.151\pm0.015\%$, $0.446\pm0.38\%$, and $0.569\pm0.014\%$ in the WFC3 bandpass, \Spitzer\ $3.6$, and \Spitzer\ $4.5\mu$m, respectively.  The best fit blackbody to the WFC3 dayside spectrum has a brightness temperature of $2930 \pm 40$ K, making WASP-103b among the hottest exoplanets ever observed.}
	\item{The phase curves have large amplitudes ($0.8 -0.9\times$ the secondary eclipse depth), and small offsets in peak brightness from the substellar point (consistent with zero degrees at all wavelengths). These characteristics indicate inefficient redistribution of heat to the nightside, as seen in other very hot Jupiters \citep{komacek17}.} 
	\item{We fit the phase variation with the \texttt{spiderman} package \citep{louden17} to evaluate different models of the planet's climate, including a two-temperature map, a physically-motivated kinematic map, and spherical harmonics. The spherical harmonic temperature map generally provides the best fit to the data; however, all the maps produce reasonable fits ($\chi^2_\nu$ near unity), and there are large differences in temperature between them (up to 1000 K at a given latitude/longitude). Breaking the degeneracy between different climate maps will require higher precision phase curves and/or secondary eclipse mapping \citep[e.g.][]{dewit12}.}
\item{We calculated phase-resolved spectra in ten orbital phase bins. The \HST/WFC3 spectra are consistent with  blackbody emission from the planet at all orbital phases. The best fit brightness temperatures ranges from $1880\pm40$ K (phase $\phi = 0.1$) to $2930\pm40$ K on the dayside. We attribute the absence of water features at WFC3 wavelengths to (1) H$_2$O dissociation on the dayside and (2) additional near-IR opacity from H-, TiO/VO and FeH.}
\item{The \Spitzer\ data are \emph{not} consistent with the best fit blackbody to the WFC3 data: they have a higher brightness temperature at phases $\phi = 0.2 - 0.5$, which transitions to a lower brightness temperature on the nightside ($\phi = -0.2 - 0.1$). An atmospheric retrieval analysis suggests that these characteristics are likely due to CO features in the infrared and a temperature inversion on the dayside but not the nightside.}
\item{The measured transmission spectrum is featureless (after correcting for nightside emission from the planet). 3D model predict water features in transmission that could be detected with future high precision observations.}
	\item{We characterized the composition with a 1D grid-based retrieval that assumes thermochemical and radiative-convective equilibrium. The atmosphere is moderately metal-enriched ($23^{+29}_{-13}\times$ solar; and $>1\times$ solar at $3\,\sigma$ confidence). This value is somewhat higher than what is observed for other gas giants \citep[e.g.][]{wong04, kreidberg14b}, but may be indicative of intrinsic scatter in the relationship between atmospheric metallicity and planet mass predicted by theoretical models \citep{fortney13, mordasini16}. However, the metallicity is strongly sensitive to the $4.5\,\mu$m \Spitzer\ eclipse depth, and additional observations would be useful in confirming the metal enhancement.  In addition to metallicity, we also infer an upper limit on the carbon-to-oxygen ratio of 0.9 ($3\,\sigma$ confidence). This estimate agrees with expectations from planet formation models that pollution from water ice in planetesimals leads to $\mathrm{C/O} < 1$ in gas giant atmospheres \citep{mordasini16, espinoza17}. The best fit temperature pressure profile has a thermal inversion from $\sim10^{-2} - 10^{-3}$ bars due to TiO/VO absorption at high altitudes.} 
	\item{We ran several 3D GCMs to compare to the data, including a nominal model with a cloud-free, solar composition, a metal-enriched model ([Fe/H] = 0.5), and two models with Lorentz drag. The GCM with a Lorentz drag timescale of $10^4$ s matches the data best. This model has an equator-to-pole wind pattern that reproduces the small observed hotspot offsets and large phase curve amplitudes. We made a simple order-of-magnitude estimate for the magnetic field strength required to produce this fast drag timescale, and found it implies a magnetic field of $\sim1$ Gauss.}
	\item{We compared the spectra of WASP-103b at phases 0.5 (dayside), 0.25 (quadrature), and 0.1 (nightside) to brown dwarfs and directly imaged companions of similar temperature. We quantify the strength of the water feature and find that both brown dwarfs and imaged companions show evidence for water absorption at $1.4\,\mu$m, whereas the WASP-103b dayside and quadrature spectra do not. We attribute the difference to two factors: WASP-103b's irradiation environment, which changes the temperature pressure profile, and its low surface gravity, which pushes the photosphere to higher altitudes where water dissociates more easily. The WASP-103b nightside spectra have larger uncertainties and are consistent with the water feature amplitudes for other objects; higher precision phase curves are needed to detect water on the nightside.}
\end{itemize}

These results provide a first look at the global composition and thermal structure of WASP-103b.  The planet is complex, with changes in temperature profile with longitude, possible gradients in composition from dayside to nightside, and circulation patterns that may be influenced by the magnetic field. These findings highlight the 3D nature of exoplanets and illustrate the importance of phase curve observations to develop a comprehensive understanding of their atmospheric chemistry and physics.  


\acknowledgments
Support for \textit{HST} program GO-15050 was provided by NASA through a grant
from the Space Telescope Science Institute, which is operated by the Association
of Universities for Research in Astronomy, Inc., under NASA contract NAS
5-26555. Support for \textit{Spitzer} program 11099 was provided by NASA through
an award issued by JPL/Caltech.  The data presented in this paper were obtained
from the Mikulski Archive for Space Telescopes (MAST) and the NASA/IPAC Infrared
Science Archive. The Infrared Science Archive is operated by the Jet Propulsion
Laboratory, California Institute of Technology, under contract with the National
Aeronautics and Space Administration.  This work also made use of the Python
packages  SciPy and NumPy \citep{jones_scipy_2001, van2011numpy}. The authors
are grateful for helpful conversations with Caroline Morley, Ming
Zhao, Kimberly Cartier, Hannah Diamond-Lowe, and Nick Cowan. We also thank the
organizers of the 2016 Santa Cruz Kavli Summer Program and the 2017 Ringberg
Atmospheres of Disks and Planets meeting for facilitating productive discussion
and collaboration.  J.L.B. acknowledges support from the David and Lucile
Packard Foundation.  J.M.D. acknowledges support from the European Research
Council (ERC) under the programme Exo-Atmos (grant agreement no. 679633). G.W.H.
and M.H.W. also acknowledge long-term support from Tennessee State University
and the State of Tennessee through its Centers of Excellence Program. M.R.L.
acknowledges NASA XRP grant NNX17AB56G for partial support of the theoretical
interpretation of the data, as well as the ASU Research Computing staff for
support with the Saguaro and Agave compute clusters. Finally, we thank the
anonymous referee for a thoughtful and detailed report that improved the quality of the
paper.

 \newcommand{\noop}[1]{}

\end{document}